\documentclass{aa}  
\usepackage{graphicx}
\usepackage{txfonts}
\usepackage{natbib}
\bibpunct{(}{)}{;}{a}{}{,}
\pdfoutput=1 
\newcommand{\doceCO}{\mbox{$^{12}$CO}}
\newcommand{\doce}{\mbox{$^{12}$CO}}
\newcommand{\dsiete}{\mbox{C$^{17}$O}}
\newcommand{\docho}{\mbox{C$^{18}$O}}

\newcommand{\gohab}{\mbox{GoHam}}
\newcommand{\goham}{\mbox{Gomez's Hamburger}}
\newcommand{\trece}{\mbox{$^{13}$CO}}
\newcommand{\treceCO}{\mbox{$^{13}$CO}}
\newcommand{\jsc}{\mbox{$J$=6$-$5}}
\newcommand{\jtd}{\mbox{$J$=3$-$2}}
\newcommand{\jdu}{\mbox{$J$=2$-$1}}

\newcommand{\kms}{\mbox{km\,s$^{-1}$}}

\newcommand{\ms}{\mbox{$M_{\mbox{\sun}}$}}
\newcommand{\ls}{\mbox{$L_{\mbox{\sun}}$}}

\newcommand{\lsim}{\raisebox{-.4ex}{$\stackrel{\sf <}{\scriptstyle\sf \sim}$}}
\newcommand{\gsim}{\raisebox{-.4ex}{$\stackrel{\sf >}{\scriptstyle\sf \sim}$}}

\newcommand{\farcss}{\mbox{\rlap{.}$''$}}
\begin{document}
   \title{The physical conditions in Gomez's Hamburger
   (IRAS\,18059-3211), a pre-MS rotating disk  
}

   \author{
          V. Bujarrabal\inst{1}
	  \and
          K. Young
\inst{2}
          \and
	A. Castro-Carrizo
\inst{3}
          }

   \offprints{V. Bujarrabal}

   \institute{             Observatorio Astron\'omico Nacional (OAN-IGN),
              Apartado 112, E-28803 Alcal\'a de Henares, Spain\\
              \email{v.bujarrabal@oan.es}
\and
  Harvard-Smithsonian Center for Astrophysics, 60 Garden Street,
              Cambridge, MA 02138, USA                 \\
              \email{rtm@cfa.harvard.edu}
              \and
 Institut de RadioAstronomie Millim\'etrique (IRAM), 300 rue de la
              Piscine, 38406 St.\ Martin d'H\`eres, France    \\
              \email{ccarrizo@iram.fr}
           }

   \date{Accepted }

  \abstract 
{} 
  {We aim to study the structure, dynamics and physical
  conditions of Gomez's Hamburger (IRAS\,18059-3211; \gohab). 
  We confirm that \gohab\ essentially consists of a flaring disk
  in keplerian rotation around a young, probably pre-MS star.}
{We present high resolution SMA maps of \doce\ \jdu, \trece\ \jdu,
  \doce\ \jtd, and \dsiete\ \jtd, as well as data on \doce\ \jsc\ and
  the continuum flux at these wavelengths. Spatial resolutions up to
  1$''$ are obtained. Except for the \dsiete\ data, the dynamical
  ranges are larger than 10. The maps are compared to a numerical
  model, which simulates the emission of a rotating disk with 
  the expected general properties of such objects; a very
  satisfactory fitting of our maps is obtained. The meaning and
  reliability of our results are thoroughly discussed. 
}
{Our observations allow measurement of the main properties of \gohab\ at
  scales between $\sim$ 1$''$ ($\sim$ 5 10$^{15}$ cm, for the assumed
  distance, 300 pc) and the total extent of the nebula, 14$''$. We are
  able to measure the global structure of the gas-rich disk, which is
  found to be flaring, and its dynamics, which is clearly dominated by
  keplerian rotation, with a very small degree of turbulence. The
  combination of different lines, particularly showing different
  opacities, allows us to reasonably estimate the distributions of the
  gas temperature and density. We clearly find a significant and
  sharp increase in temperature at large distances from the
  equator, accompanied by a decrease in density of the same
  order. Finally, we identify a condensation in the southern part of
  the disk that has no counterparts in the rest of the nebula. This
  condensation is quite extended (about 5 10$^{15}$ cm), contains a
  significant amount of mass (roughly, $\sim$ 6 10$^{-3}$ \ms), and
  seems to be associated with a detectable distortion of the global
  rotation kinematics. We discuss several possible interpretations of
  that feature.}
{}
\maketitle

\section{Introduction}

Gomez's Hamburger (IRAS\,18059-3211; hereafter \gohab) is a very
interesting nebula, originally identified on a plate by A.\ G\'omez in
1985, that still remains poorly studied. It shows a
spectacular optical image \citep[see][and HST images in press release
number 2002-19]{ruizetal87,buj08}, in which a dark lane of dust
separates two flat, bright regions, presumably illuminated by a central
star that remains hidden by the equatorial disk.  The central
star has been classified as an A0-type star, from spectroscopic
analysis of the scattered light \citep{ruizetal87}. \gohab 's spectral energy
distribution (SED) shows two maxima, in the optical and FIR
\citep{ruizetal87}, which respectively correspond to the stellar
emission scattered by the bright lobes and to radiation reemitted by
dust grains in the dark equatorial region.

Originally, \gohab\ was thought to be a post-AGB nebula
\citep{ruizetal87}. However, high-resolution CO maps
\citep[][hereafter, Paper I]{buj08} indicate that the nebula is almost
exactly in keplerian rotation, around the symmetry axis clearly
identified from its image. No component other than the rotating,
flaring disk, at least at the large scale probed by the arcsec
resolution attainable with mm-wave interferometry, seems to be present
in the nebula.  \citet{buj08} estimated the central (presumably
stellar) mass and concluded that, for any possible distance, its high
value in combination with the relatively low total luminosity of the
source is not compatible at all with the idea that \gohab\ is an evolved
object. These authors conclude that \goham\ is very probably a
pre-main-sequence object at a much smaller distance than previously
believed, at about 500 pc. The value of the distance remains however
uncertain. \citet{wood08} built a model for the dust emission and
scattering that can explain the optical and IR SED for a distance of
about 300 pc. The total disk mass determined by these authors, 0.3 \ms,
is very similar to that deduced in Paper I from a simple analysis of
the mm-wave dust emission (taking into account the different distances
assumed in these papers), and is very probably dominated by very dense
inner regions.  Recently, \citet{berne08} have also suggested a
distance of about 200-300 pc, arguing that the spectral type could be
slightly less early than that obtained by \citet{ruizetal87}. The
determination of the distance is based on the comparison of several
observational parameters obtained in \gohab\ with theoretical
evolutionary tracks and previous data for other young stars. The
relevant parameters are: the total luminosity (obtained from
integration of the SED, but dependent on the anisotropy of the
dust-processed radiation, see below)
the central mass
(from the nebula rotation), and the stellar surface temperature 
(see discussion on the stellar type accuracy by Bern\'e et al.) If
the stellar type is slightly less early than that proposed by
\citet{ruizetal87}, a distance of about 300 pc is compatible with all
existent data.  These smaller values of the distance yield a slightly
smaller linear radius for the disk, more in agreement with results
usually found in similar objects. We accordingly adopt a distance
of 300 pc.

The SED observations of \gohab\ are quite complete \citep{ruizetal87},
but the estimate of the luminosity requires to introduce significant
corrections due to the extreme viewing angle of the source and the
strong disk opacity at short wavelengths (see Paper I). For the
distance assumed here, we can expect a luminosity of $\sim$15 \ls, with
an uncertainty of about a factor 2. Comparison of this value of the
luminosity with pre-MS evolutionary tracks \citep[see][Paper
I]{ancker98} indicates a stellar mass of about 2 \ms. This value
is relatively low compared to the mass derived from the disk rotation
curve, $\sim$ 3 \ms, suggesting a significant contribution to the
central mass of the innermost disk or even of a low-mass stellar
companion.  See further discussion in Paper I; the nature of the
central star(s) will be reviewed in Sect.\ 4, taking into account our
new results. 

\citet{buj08} presented only \doce\ \jdu\ map obtained with the SMA,
together with a preliminary modeling of the data. These models were
severely hampered by the limited amount of data given in that work. In
particular, the high optical depth expected in this line prevented any
accurate determination of the gas density distribution. A significant
revision of some parameters tentatively discussed by \citet{buj08} was
therefore necessary. In this paper, we also present maps of \trece\
\jdu, \doce\ \jtd, and \dsiete\ \jtd.  The combination of the \doce\
\jdu\ line (reanalyzed for this work) with the \trece\ and \dsiete\
lines, which are optically thin, and the \jtd\ emissions, requiring
significantly more excitation, has allowed a very detailed modeling of
the source. We think we can now propose realistic distributions at
large scale of the gas density and temperature in \goham, distributions
that, as we will see, appear compatible with theoretical
ideas on the properties of passive flaring disks in keplerian rotation.

   \begin{figure*}

\centering
   \rotatebox{270}{\resizebox{8cm}{!}{ 
\includegraphics{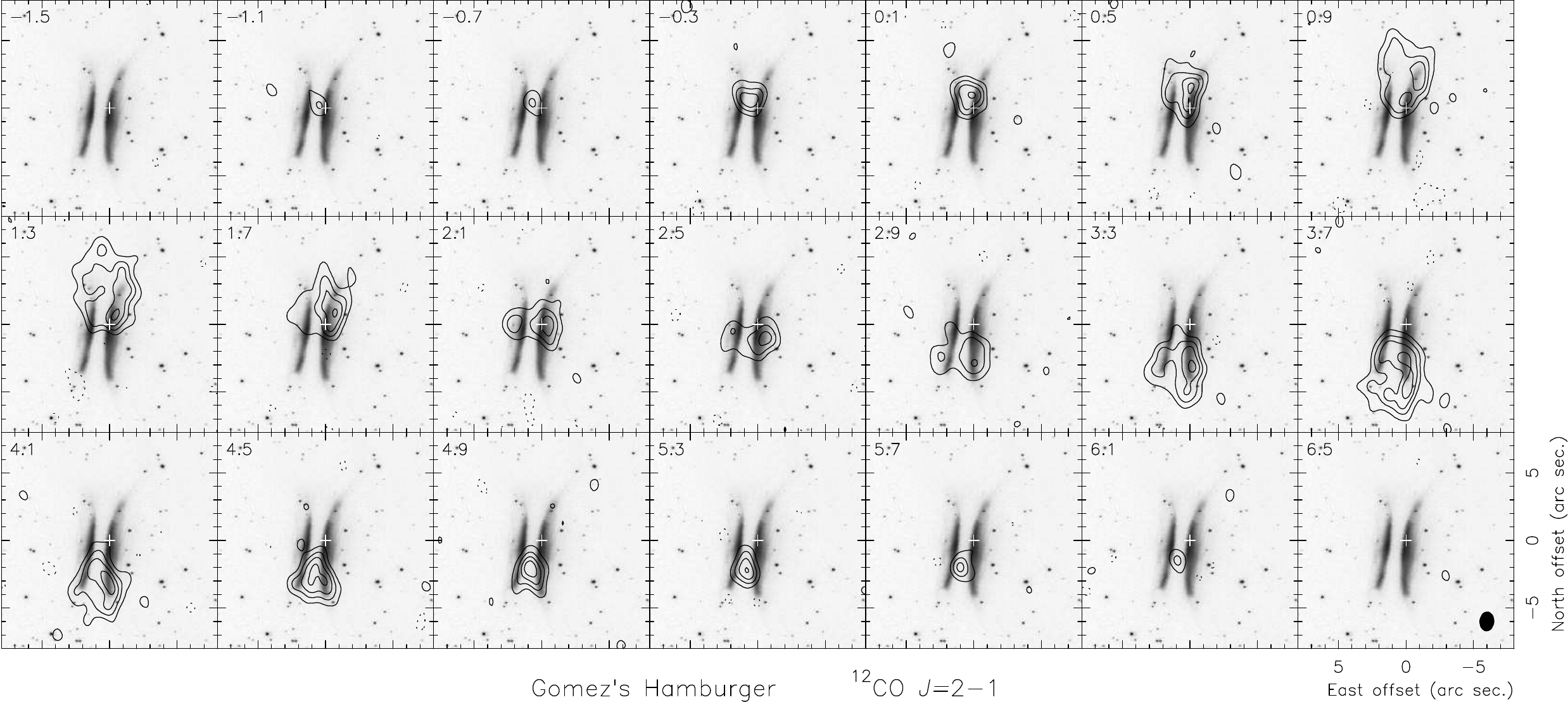}
}}
\vspace{-0cm}
   \caption{Channel maps of the \doceCO\ \jdu\ line from Gomez's
   Hamburger, continuum has been subtracted. First contour and contour
   step are 0.28 Jy/beam, approximately equal to 3$\sigma$; 
negative values are represented by dashed
   contours. The LSR velocity in \kms\ for the center of each channel
   is indicated in the upper left corner. The J2000 coordinates of the
   reference position, the cross in the maps, are R.A.\ = 18:09:13.37,
   Dec = --32:10:49.5. In the last panel we represent the synthetic beam
   at half-intensity (black ellipse). The HST image is also shown for
   comparison. See details in Sect.\ 2. 
}
              \label{maps}%
    \end{figure*}

   \begin{figure*}
\centering   \rotatebox{270}{\resizebox{8cm}{!}{
\includegraphics{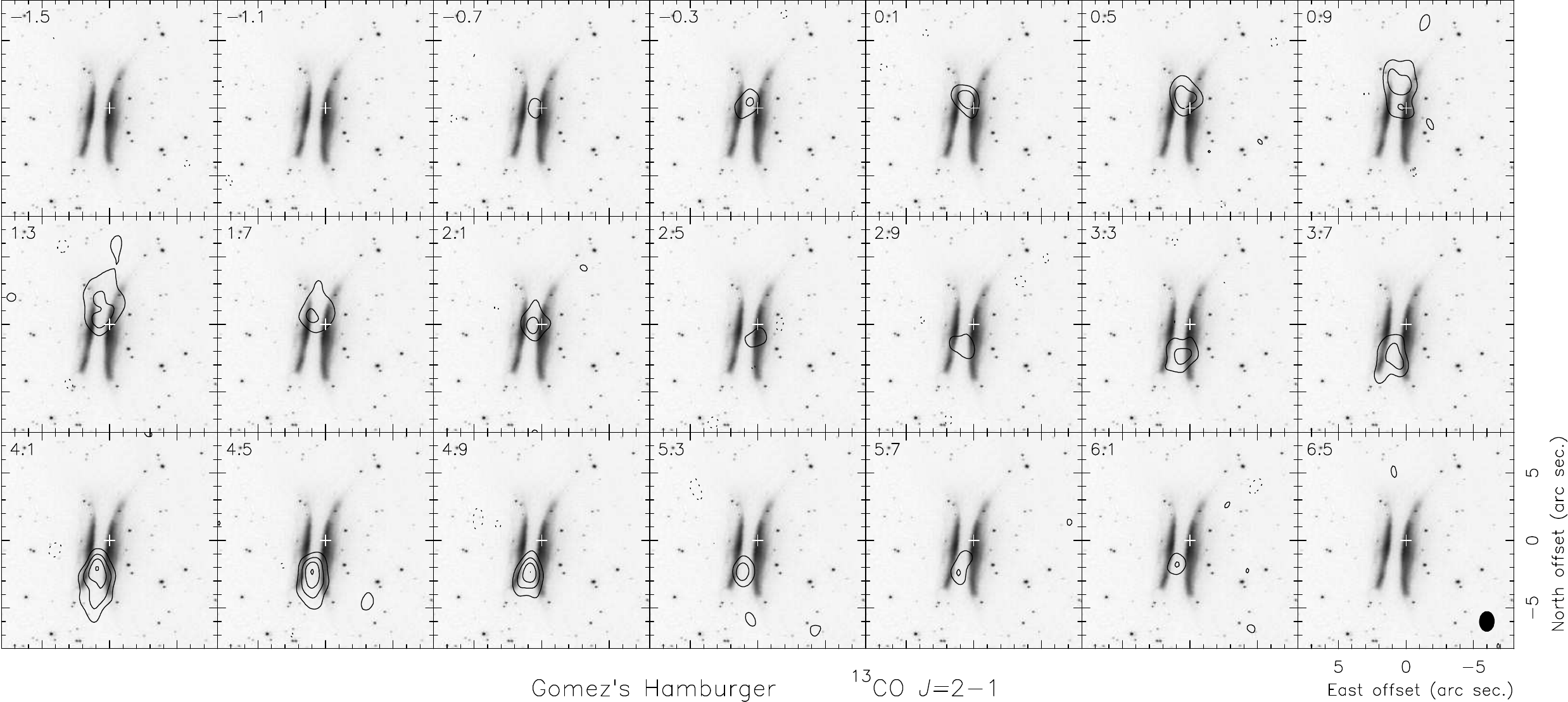}
}}
\vspace{-0cm}
   \caption{Same as Fig.\ 1, but for the channel maps of the \trece\
   \jdu\ line observed from \goham. First contour and contour
   step are also 0.28 Jy/beam.
}
              \label{maps}%
    \end{figure*}

   \begin{figure*}
\centering   {\resizebox{17cm}{!}{\includegraphics{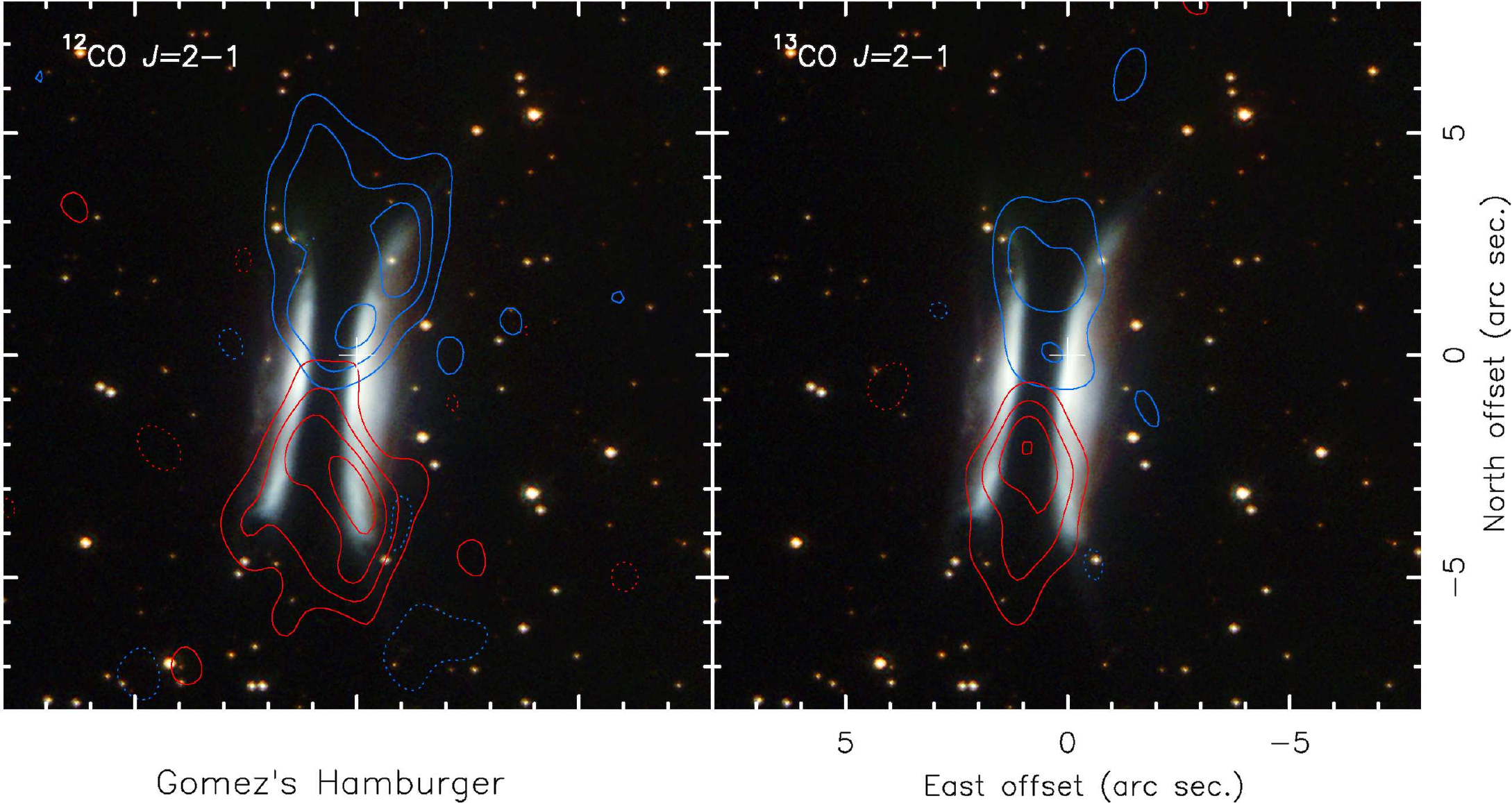}
}}
\vspace{-0cm}
   \caption{Zoomed \doce\ and \trece\ \jdu\ emission for velocities 0.9
   and 4.1 \kms, respectively blue (north) and red (south) contours,
   superposed to the HST image of \gohab; from Figs.\ 1 and 2, see
   further details in Sect.\ 2.
}
              \label{maps}%
    \end{figure*}

\section{Observations}

We present observations of the \doce\ \jdu\ (230.5 GHz), \trece\ \jdu\
(220.4 GHz), \doce\ \jtd\ (345.8 GHz), and \dsiete\ \jtd\ (337.1 GHz)
lines in \goham\ (IRAS 18059-3211, \gohab), obtained with the
Submillimeter Array (SMA). We also observed \doce\ \jsc\ (691.5 GHz),
but did not detect the line emission.
\doce\ and \trece\ \jdu\ were observed simultaneously in 2006. Details
of the observations, calibration and data reduction for these lines may
be found in \citet[][Paper I]{buj08}.  The \doce\ \jdu\ data set has been
reduced again, with respect to the observations presented in Paper I. 
The continuum flux at $\lambda$ = 1.3 mm was measured to be 0.3 Jy.

The 345 GHz and 690 GHz data were taken on the night of 2007 March 21
(UT), when the array was in the "Compact North" configuration, a low
resolution configuration with longer north-south baselines intended to
produce nearly circular synthesized beams for low declination sources.
Seven of the SMA's eight antennas were usable in the 345 GHz band, and
six were usable in the 690 GHz band.  The 345 GHz receivers were tuned
to position the \doce\ \jtd\ line at 345.796 GHz in the receivers'
upper sideband (USB) and the \dsiete\ \jtd\ line at 337.061 GHz in the
lower sideband (LSB). The 690 GHz receivers were tuned to position
\doce\ \jsc\ at 691.473 GHz in the USB.  The J2000 coordinates used for
\gohab\ were RA = 18:09:13.37, Dec = --32:10:49.5, and correspond to
the central cross in our images.  Observations began at 11:00 UT and
finished at 21:10 UT.  During that period, the precipitable water
vapor above Mauna Kea varied from 0.6 to 1.0 mm.  As the source
transited at an elevation of 38 degrees, the double sideband (DSB)
system temperature ($T_{\rm sys}$) for the seven 345 GHz receivers
ranged from 133 K to 235 K, while $T_{\rm sys}$ ranged from 1600 K to
2300 K for the 690 GHz receivers.  The correlator was configured to
have a uniform resolution of 812.5 kHz per channel across the full 2
GHz (DSB) bandpass of each receiver.  This resolution is equivalent to
0.70 \kms\ for the \jtd\ lines, and 0.35 \kms\ for the \jsc\ line.
Observations of Ganymede (16.5 degrees from the source) and the quasar
1924-292 (15.7 degrees from the source) were made every half hour for
calibration of the instrument's complex gain.  Saturn and 3C 273 were
observed before the source rose, to provide bandpass calibration.

   The data were calibrated with the MIR package.  For the 345 GHz
data, crosscorrelation observations of Saturn, 3C 273 and Ganymede were
used to derive antenna-based bandpass data to correct for the
instrument's spectral response.  Ganymede was used for flux
calibration, and the derived 850 micron flux for 1924-292 was 2.5 Jy.
There were no phase jumps in the data, and complex gain calibration was
done with a 5th order polynomial fit to the amplitude and phase of the
quasar data.  The source was tracked from an hour angle of --3.5 to
+3.5 hours, and the total on-source integration time was 4 hours.
Projected baseline lengths ranged from 9.1 to 136 kilolambda.  The
\doce\ and \dsiete\ \jtd\ spectral lines were clearly detected, as was
the 850 micron (345 GHz) continuum emission, with a total flux of 0.8
Jy.  The astrometry was checked by using the Ganymede data to phase
calibrate the 1924-292 data, and mapping 1924-292.  The map showed
1924-292 as a point source with a position error of 0.2 arc seconds.

   The 690 GHz band data were more difficult to process, in part
because there were instrumental problems.  Three of the correlator's 24
spectral bands were defective on some baselines.  One receiver lost
phase lock for 45 minutes during the track, and relocking the receiver
produced a phase jump on baselines to that antenna.  All data taken
with $T_{\rm sys}$ $>$ 6000 K (on either antenna of any baseline) were
discarded.  Finally, much of the data were taken after sunrise, when the
pointing of the array antennas is degraded.  The crosscorrelation
amplitude of Ganymede was checked on all baselines, and the source data
were discarded whenever the nearest (in time) Ganymede scans had low
amplitudes.  Altogether, roughly 1/3 of the 690 GHz data were discarded
for these reasons.  Saturn, 3C 273 and Ganymede again were used to
provide instrument bandpass calibration.  1924-292 was not detected on
short timescales in the 690 GHz data, so only Ganymede could be used
for gain calibration.  Ganymede was resolved by the array at 690 GHz,
which caused the amplitude of this calibrator to pass through zero as
the phase jumped by 180 degrees several times during the track.
Projected baselines for the on-source data ranged from 17.9 to 230
kilolambda. The 450 micron (690 GHz) continuum emission from \gohab\
was detected, with a total flux of 1.5 Jy, but CO \jsc\ line emission
was not, with a typical noise in the maps of $\sim$ 1 Jy/beam.

   Mapping was done with the NRAO AIPS package.  The synthesized beam
shapes were 3\farcss07 $\times$ 1\farcs35 with major axis position
angle, PA = 37.4$^\circ$ for the \dsiete\ \jtd\ maps, 2\farcss 95
$\times$ 1\farcss31 with PA = 37.6$^\circ$ for the \doce\ \jtd\ maps,
and 3\farcss59 $\times$ 0\farcss82 with PA = --5.7$^\circ$ for the 450
micron continuum map.

The resulting maps per velocity channel of the four detected lines are
shown in Figs.\ 1 to 5. 

In Figs.\ 1 and 2 we also represent the HST image of \gohab, to allow a
better comparison with the CO emission and between both \doce\ and
\trece\ \jdu\ lines. In Fig.\ 3 we show a zoom of the data from Figs.\
1 and 2 for the emission at velocities 0.9 and 4.1 \kms\ LSR, expected
to represent the emission from regions closer to the plane of the
sky. To show the HST images we have chosen the beautiful color image
from press release 2002-19 provided by the HST newscenter (obtained
from WFPC2 images, filters: F675W, F555W, F450W), see
http://hubblesite.org/newscenter/. The coordinates were derived from
the NICMOS images in the HST archive (taken on April 12, 2006, HST
project 10603, P.I.: D.\ Padgett; NICMOS images were more directly
comparable to the press release image than the archive WFPC2 data).  We
have made use of the Aladin viewer and database.  We recall that these
images have not been yet published in the specialized literature and
that a deep analysis of the HST imaging of \gohab\ is out of the scope
of this paper.

   \begin{figure*}
\centering\rotatebox{270}{\resizebox{5.6cm}{!}{ 
\includegraphics{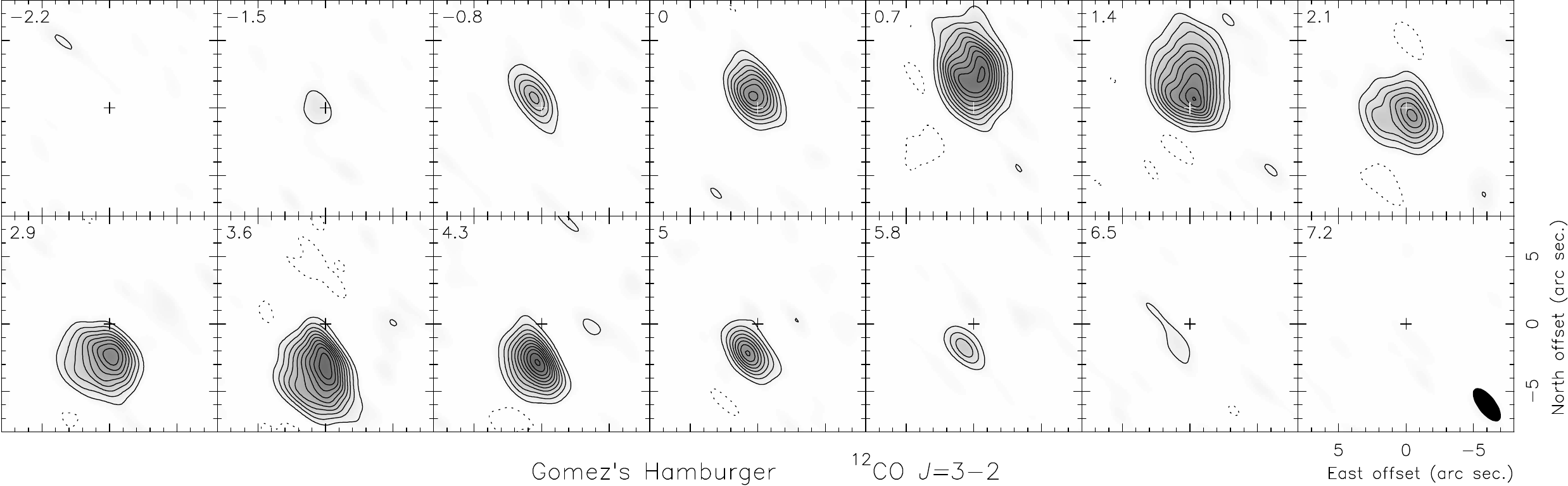}
}}
   \caption{Same as Fig.\ 1, but for the channel maps of the \doce\
   \jtd\ line observed from \goham. First contour and contour
   step are 0.5 Jy/beam. Note that the spectral and spatial
   resolutions of these observations are somewhat poorer.}
              \label{maps}%
    \end{figure*}

\section{CO emission model}

In order to extract as much information as possible from our maps, we
have used a code that simulates the emission from a rotating model
disk. Our model provides brightness distributions for LSR velocity
channels, to be directly compared with the observations.  We assume a
shape of the CO cloud and a spatial distribution of the velocity,
temperature, density, and CO relative abundance. Given these
parameters, our code calculates the emissivity of the lines at each
point of the disk, and the brightness for a number of lines of sight
solving the full radiative transfer equation. Opacity effects and
velocity shifts are accurately taken into account. Such a brightness
distribution is convolved with the synthetic beam, and images
with the same units as the observed ones are produced.

The code itself is similar to that described by \citet{buj08}, Paper
I. See further details there, including discussions on the basic
assumptions. The method used to find the best fit and the uncertainties
in the derived parameter values are presented in the Appendix.

However, the disk model used here is much more complex than that in 
Paper I, mainly because the large amount of empirical data presented now
allows us to undertake an accurate description of the structure and
physical conditions in the disk. We have tried to reproduce all our SMA
maps from the model predictions and, at the same time, to keep our disk
model compatible as much as possible with theoretical ideas on the
properties of rotating disks around young stars. As we will see
in the next subsection, the model has become relatively complex, but
the different disk components always correspond to features actually
identified in our data.

   \begin{figure*}
   \centering \rotatebox{270}{\resizebox{5.6cm}{!}{ 
\includegraphics{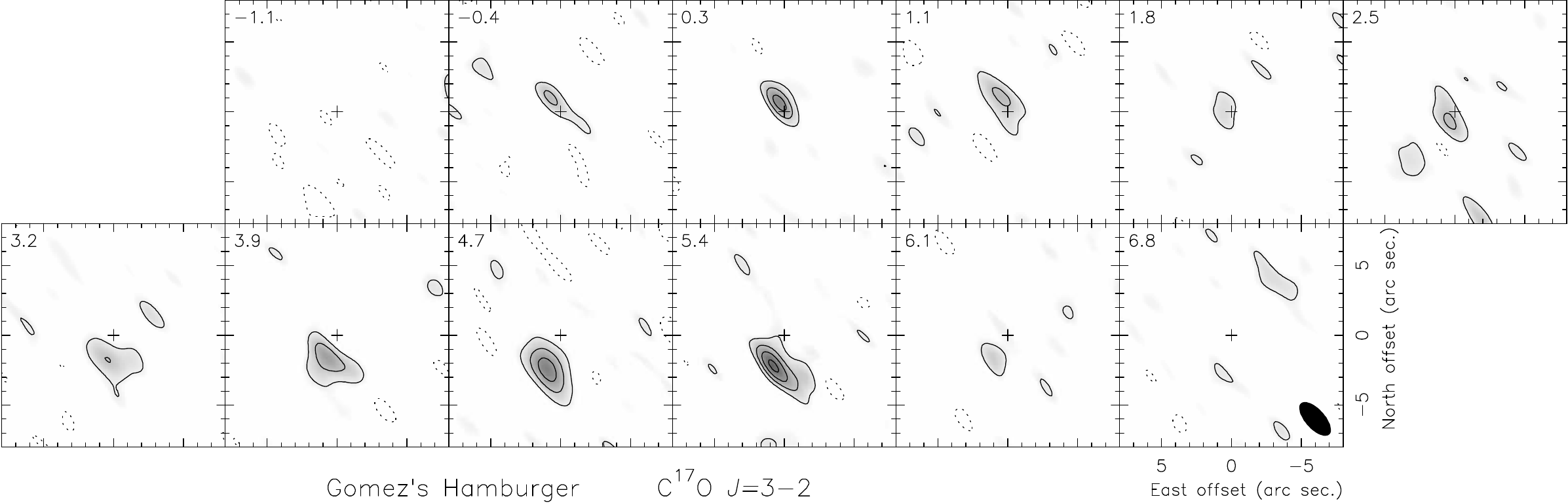}
}}
   \caption{Same as Fig.\ 3, but for the channel maps of the \dsiete\
   \jtd\ line observed from \goham.  First contour and contour step are
   0.3 Jy/beam.}
              \label{maps}%
    \end{figure*}

\subsection{Disk shape and dynamics}

From a first inspection of our maps and the HST image (Figs.\ 1 to
3), it is clear that the nebula is essentially a rotating disk seen
almost edge on, with a significant axial symmetry and almost exactly
keplerian dynamics. We can also see that the disk is roughly flaring,
being significantly wider at larger distances from the center. The disk
symmetry axis is almost placed in the plane of the sky, slightly tilted
such that the western part of the disk points towards us. The
projection of the disk symmetry axis on the sky plane is almost in the
east-west direction, with position angle, PA $\sim$ 85$^{\circ}$, i.e.\
the disk appears almost parallel to the declination axis.

We accordingly assume that the emission comes from a flaring disk
in keplerian rotation and showing axial symmetry. Our calculations 
confirm that the observations can be explained under these
assumptions. Two dimensions are then enough, $r$, the distance to the
disk axis, and $z$, the distance to the disk equator.

We assume the simplest flaring geometry, with a disk width just
proportional to $r$. Note that, in our case, the disk boundaries
represent the region where CO is still abundant, before it is
significantly photodissociated (due to the UV radiation from the
central star or from the galactic background), although the comparison
of the \doce\ maps with the HST scattered light images (Figs.\ 1 to 3)
suggests that the \doce\ disk in fact occupies the whole nebula.
To keep the model as simple as possible, we assume constant CO
abundance, $X$(CO), and isotopic ratios. Note that the CO abundances
and density laws are not independent parameters in our model;
therefore, the value of $X$(\doce) given here is in fact an assumption
chosen to be reasonably in agreement with expectations. We recall
that the molecular abundances in inner regions of the disk may be
significantly smaller than the usual ones in the interstellar medium,
mainly due to depletion onto grains \citep[see discussion in e.g.][and
references therein]{panic08,thi04}. As mentioned above, in the outer
regions of the disk we could expect significant dissociation of CO. We must
therefore keep in mind that, although the CO abundance assumed here for
\doce, 10$^{-4}$, is moderate, its value could be still lower in
the innermost regions.

As we can see in our \doce\ \jdu\ maps, particularly at 1.3 \kms\ LSR,
the shape of the outer part of the disk seems to be rounded, instead of
flaring (at large distances, $r$ $>$ $R_{\rm m}$, to the center). The
model predictions are very clearly different from the observations if
we assume that the flaring geometry continues up to the outer disk
radius. This assumption is compatible with calculations of the shape of
the CO abundance variations at large values of $r$, see e.g.\
\citet{jonkheid06,jonkheid07}. We will see from our model calculations
how this disk shape well reproduces the observations. The shape of the
disk is then given by the following parameters: The outer maximum and
inner minimum disk radii, $R_{\rm out}$ and $R_{\rm in}$; the
intermediate radius up to which the disk is flaring, $R_{\rm m}$; and
the width at a given value of $r$, $H$($r_0$), such that $R_{\rm in}$
$\leq$ $r_0$ $\leq$ $R_{\rm m}$. Between $R_{\rm m}$ and $R_{\rm out}$,
the disk width is assumed to vary following an elliptical function. See
the resulting disk shape in Fig.\ 7.

   \begin{figure*}
   \centering \rotatebox{270}{\resizebox{8cm}{!}{ 
\includegraphics{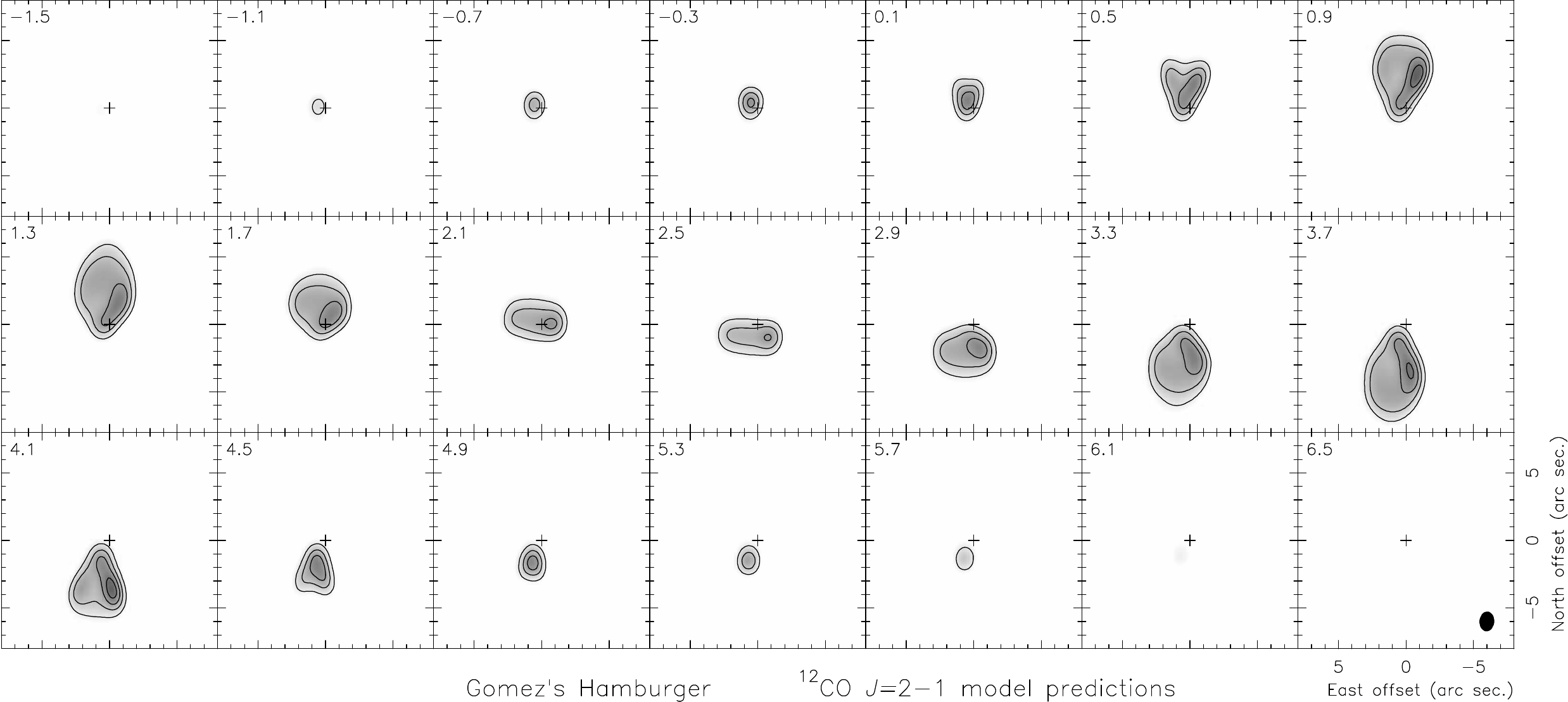}
}}
   \caption{Predictions of the \doceCO\ \jdu\ line brightness from our
   best model fitting for the \gohab\ disk. The spatial scale and 
   contours are the same as in the observations, Fig.\ 1.
}
              \label{maps}%
    \end{figure*}

The kinematics is
assumed to be given by the rotational velocity,
following a keplerian law with $r$, i.e.\ varying proportionally to
1/$\sqrt{r}$. As in paper I, we find that some amount of local
turbulent dispersion helps to fit the data, but that its value must be 
kept very small. We then assume the existence of a constant local
dispersion $V_{\rm turb}$.

\subsection{Density and temperature}

Direct inspection of our data suggests that the density and temperature
distribution must follow relatively complex laws, that we will
introduce in our model, trying to keep them compatible with general
theoretical ideas.

We can see, first of all, a remarkable difference between the \doce\
and \trece\ \jdu\ maps, see Figs.\ 1 and 2 and a zoom of
intermediate velocities in Fig.\ 3. The \trece\ brightness
distribution is significantly narrower (less extended in the direction
of the disk axis, which is almost equivalent to the east-west
direction) and shorter (less extended in the disk plane
direction). Nevertheless, in regions in which both lines are intense
(in the disk plane, slightly northwards and southwards from the star),
the \doce\ \jdu\ brightness is only slightly higher. On the other hand,
in the disk edges, where only \doce\ emission appears, the brightness
of this line is often significantly higher than in the central regions
(where both \jdu\ transitions are detected).

Since \trece\ emission is certainly expected to be much less optically
thick than \doce\ emission and both lines are easily thermalized, this
different behavior between them obviously means that there are wide
outer edges of the disk in which the density significantly decreases
and the temperature increases. These edges are both more or less
parallel to the disk equator (appearing from certain values of
$\mid$$z$$\mid$) and perpendicular to it (appearing from certain values
of $r$). This behavior is very noticeable in the direction
perpendicular to the plane, i.e.\ sharp variations of density and
temperature take place from a certain distance to the disk
equator, and are particularly well identified in the maps. We can also
see in our \doce\ \jdu\ maps that one of these edges parallel to the
equator (so, almost parallel to the north-south direction), the
western one, is significantly more intense than the other. This is
due to a simple radiative transfer effect: the emission from the
eastern edge is absorbed by the cooler and opaque inner regions,
which does not occur for the emission from the other side, which is
closer to us. As we will see, our model reproduces very convincingly
this phenomenon. Of course, nothing comparable is seen in the optically
thinner \trece\ emission, which, because of the low density of the hot
edges, is only intense in the dense central regions.

Therefore, to reproduce the observations of the different lines, we have
to introduce several components in our density and temperature laws. At
a minimum we need a cool dense region close to the equator, plus two
edges with significantly less density and higher temperature (which are
more or less parallel to the equator and separated quite sharply from
the previous equatorial region), plus an outer region far from the
disk axis with similar but smaller variations in these parameters.

We finally introduce a new correction: to
explain the relatively intense emission for regions close to the
central star, in spite of their small volume in a flaring disk, the
density and temperature must increase from the standard values at
small enough distances from the star. As we will discuss in the
Appendix, the physical properties of these regions, unresolved in our
observations, are not well determined. 

The equatorial density and temperature ($n_{\rm eq}$, $T_{\rm eq}$) are
supposed to follow potential laws, $n_{\rm eq}(r) \propto
1/r^{\alpha_n}$ and $T_{\rm eq}(r) \propto 1/r^{\alpha_T}$. Such laws
are often found to be compatible with studies of molecular gas in disks
around young stars \citep[e.g.][]{dutrey07}, and we find that these
simple laws are enough for the amount of information actually contained
in our data. In practice, we assume that $\alpha_n$ is not a free
parameter but equal to 1, a value that can fit our results. Note that
the assumption of constant accretion rate, sometimes assumed to derive
the density law, can hardly be justified in a relatively evolved object
like \gohab, whose central star is barely accreting material.

The variation of the density with $z$, the distance to the equator is
assumed to be given by the vertical equation of hydrostatics \citep[see
  e.g.][]{dullemond07}:
\begin{equation}
\frac{{\rm d}P}{{\rm d}z} = -\rho \, \Omega^2 z ~~.
\end{equation}
Where the pressure is $P$ = $k T / m_p$, $m_p$ is the average weight of
the gas particles, $\rho$ is the density, and $\Omega(r)$ is the
angular velocity.

For a constant temperature with $z$, $T(r,z)$ = $T_{\rm eq}(r)$,
$n(r,z)$ is given by:
\begin{equation}
n(r,z) = n_{\rm eq}(r) ~ e^{-(z^2 \frac{C_1}{2 \, T_{\rm eq}(r)})} ~~.
\end{equation}
Where $C_1$ = $\Omega^2(r) \, m_p / k$ is a function of $r$, not of $z$.

To  implement the  above  mentioned sharp  change  in temperature  and
density from a  value of $\mid$$z$$\mid$, we assume  that when $n(r,z)$
reaches a certain fraction of  $n_{\rm eq}(r)$ the temperature changes by
a factor, $F_T$.  The value of $F_T$ is  obtained from comparison
between  the observations and the  model predictions, see Table 1 and
Sect.\ 4, as well as the
fraction of the density at which the jump must appear (1/10 was found to
yield acceptable model results).

We are assuming that there is also a jump of the density at this point,
by a factor $F_n$. $F_n$ can be calculated from the equation of
hydrostatics, which can be written in the form:
\begin{equation}
\frac{1}{\rho} \frac{{\rm d}P}{{\rm d}z} = -\frac{z\,C_1}{T} 
-\frac{1}{T} \frac{{\rm d}T}{{\rm d}z} ~~.
\end{equation}
For a very steep jump in T:
\begin{equation}
\frac{1}{\rho} \frac{{\rm d}P}{{\rm d}z} \sim 
-\frac{1}{T} \frac{{\rm d}T}{{\rm d}z} ~~.
\end{equation}
Whose solution, for a jump between $T_0$ and $T_1$, is: 
\begin{equation}
\frac{\rho_1}{\rho_0} \sim \frac{T_0}{T_1} ~~.
\end{equation}

So, if the temperature varies at a given value of $z$, $z_j$, by a
factor $F_T$, the density must vary at the same point by a factor $F_n$
= 1/$F_T$. Beyond this point, for $z$ $\geq$ $z_j$, we assume that the
temperature remains constant, and then the density must continue the
law already deduced for constant temperature:
\begin{equation}
n(r,z) = n(r,z_j) ~ e^{-[(z-z_j)^2 \frac{C_1}{2 \, T(r,z_j)}]} ~~.
\end{equation}
Where $z > z_j$ and $n(r,z_j)$ and $T(r,z_j)$ denote the density and
temperature just beyond the jump at $z_j$.

We note that, for the law we are assuming for $n_{\rm eq}(r)$ ($n_{\rm
eq}(r)$ $\propto$ 1/$r$), the condition for the jump we are imposing,
i.e.\ that $n(r,z)$/$n_{\rm eq}(r)$ is smaller than a certain constant
factor, is roughly equivalent to assume that the jump appears for a
fringe roughly parallel to the outer limit to the CO-rich disk in
$z$. This is reasonable, since both the photodissociation of CO and the
gas overheating are expected to be physically due to the effects of the
outer radiation field.

We have also mentioned that $T$ and $n$ must also depart from our
standard law in the outer disk, for $R_{\rm m}$ $<$ $r$ $<$ $R_{\rm
out}$; again, $n$ must show a decrease and $T$ must increase to explain
the \doce\ and \trece\ \jdu\ emission from regions far from the
symmetry axis. The $n$ and $T$ variations are assumed to be not related
in this case. See the finally adopted laws in Table 1 and Fig.\ 7. Note
that the required change in $T$ is small, but $n$ must strongly
decrease to explain the lack of \trece\ emission (and the decrease of
the \doce\ \jdu\ intensity close to $R_{\rm out}$). In any case, the
accuracy of the density value determined at the longest distances from
the axis is poor, since it mostly depends on the intensity of only one
line, which is (moderately) optically thick; see Appendix. Note
that, for the adopted physical conditions, the density in regions with
$r$ slightly larger than $R_{\rm m}$ is high enough to allow detectable
emission of \trece\ and \doce\ \jdu\ for some LSR velocities. The
central velocities of our \doce\ \jdu\ maps are affected even by the
outermost parts of the disk.

\subsection{Other parameters}

The inclination of the rotation axis with respect to the plane of the
sky must be small, as suggested by the optical and CO images. The
projection of this axis on the plane of the sky is obviously placed at
position angle, PA, slightly smaller than 90$^\circ$. We note that both
parameters may in fact vary with $r$, since such a big disk may be
warped at some degree. This seems to be the case if we compare the
inclinations in the CO maps for $V_{LSR}$ $\sim$ 1.3--1.7 \kms\ and for
0.9 or 3.3 \kms. It is probable that some regions of the disk rotate
around an axis (in projection) at PA $\sim$ 80$^\circ$, while other
regions rotate around PA $\sim$ 90$^\circ$.

The distance has been discussed in Paper I and Sect.\ 1, where the
uncertainties of its value were stressed. We adopt a distance
of 300 pc.

The systemic velocity is determined from the maps of our CO lines, it
is quite accurately measured, with an accuracy better than a 10\%. The
given value, Table 1, is compatible with all the observations presented
here.

\subsection{Fitting procedure and results}

We have chosen the best set of parameters by direct comparison of the
observed and synthetic images. We think that the comparison in the sky
plane, and not in the transformed $uv$ plane, is possible in our case
because of the relatively large extent of the source. On the other
hand, the large number of parameters, needed to explain the various
components suggested by the observational features, and the different
significance of them, prevents any systematic, blind fitting. See
detailed discussion in App.\ A.1. The \doce\ and \trece\ \jdu\ images,
which contain a lot of information, have been the basic data for this
comparison. 

In the Appendix we can also see in detail the criteria followed to
choose the acceptable models. The residual images, i.e.\ the observed
minus theoretical images, must present an r.m.s.\ noise in regions
where emission appears smaller than $\sim$ 1.5 times that found in
nearby regions (that just gives the noise of the observations). Also,
the residual images must show contours smaller than $\sim$3 times the
noise level or $\sim$2 times systematically. For instance, for a model
to be acceptable, we must not see in the resulting \doce\ \jdu\
residuals two positive or negative contours;
note that the observed \doce\ \jdu\ image do show one contour features
just due to the observational noise. For an acceptable model, the
regions in the residual images in which emission was present appear,
therefore, only slightly different from those in which there was no
emission.

We have discussed in A.2 the uncertainty in the main parameters of the
model, given by the values of each parameter that, while the others
remain unchanged, lead to results that clearly do not satisfy these
conditions.

The conditions for the model predictions to be acceptable are somewhat
relaxed for the relatively strong emission detected at about 4.5--5.5
\kms, which has no counterpart in the equivalent blueshifted
emission. We think that this emission excess is due to a condensation
in the rotating gas (Sect.\ 4 and App.\ A.3). Our model presents axial
symmetry and cannot reproduce this emission excess; we have preferred
to fit mostly the emission in the other velocities and to discuss
separately this intense clump (Sect.\ 4.1).

We present in Fig.\ 6 and in the Appendix some examples of the
resulting model maps and residuals.  Table 1 and Fig.\ 7 present
the values of the fitted parameters.

\bigskip
\begin{table*}[bthp]
\caption{Structure and physical conditions in the molecular disk in
  Gomez's Hamburger, derived from our model fitting of several CO
  rotational lines. Dependence on the assumed distance is given in the
  relevant cases.  Other parameters of the modeling are also given.}
\begin{center}                                          
\begin{tabular}{|l|lc|l|}
\hline\hline
& & & \\
{\bf Parameter}  & {\bf Law} & {\bf Values} & comments \\ 
& & & \\
\hline\hline
   &  &  &  \\
Outer radius  &   &  $R_{\rm out}$ = 3.1 10$^{16}$ ($\frac{D({\rm pc})}{300}$)
 cm &   \\
   &  &  & \\
\hline
   &  &  &  \\
Intermediate radius & & $R_{\rm m}$ = 1.7 10$^{16}$ ($\frac{D({\rm pc})}{300}$)
 cm & \\
   &  &  & \\
\hline
   &  &  &  \\
Inner radius  &  &  $R_{\rm in}$ = 3 10$^{15}$ ($\frac{D({\rm pc})}{300}$)
 cm &   \\
   &  &  & \\
\hline
   &  &  &  \\
Disk thickness & linear & $H$(10$^{16}$ cm) = 9.5 10$^{15}$ cm & \\
   &  &  &  \\
\hline
   &  &  & \\
Tangential & $V_{\rm t} \propto 1/\sqrt{r}$ 
& $V_{\rm t}$(10$^{16}$$\frac{D({\rm pc})}{300}$ cm)
= 2.1 \kms & \\
velocity  & (keplerian)~~~ & central mass: 3 ($\frac{D({\rm
 pc})}{300}$) \ms &  \\
   &  &  &  \\
\hline
   &  &  &  \\
 Temperature  & $T_{\rm eq}$ $\propto$ $1/r^{\alpha_T}$ & $T_{\rm
   eq}$(10$^{16}$$\frac{D({\rm
 pc})}{300}$ cm) = 16 K &  \\ 
(equator)  & & $\alpha_T$ = 0.3 & + increase in the edges, by a factor 3    \\
 & & & + increase beyond $R_{\rm m}$, $T$($R_{\rm out}$) = 17 K  \\
see text and Fig.\ 7 
 & & & + increase within 2 $\times$ $R_{\rm in}$, by a factor of 2   \\
  &  &  &  \\
\hline
  &  &  &  \\
 Gas density & $n_{\rm eq} \propto 1/r^{\alpha_n}$ &
$n_{\rm eq}$(10$^{16}$$\frac{D({\rm
 pc})}{300}$ cm) = 1.5 10$^6$ ($\frac{300}{D({\rm pc})}$) cm$^{-3}$
& \\
(equator)  &  & $\alpha_n$ = 1 &  + decrease with $z$, following hydrostatic eq.\  \\   
 & & & 
+ decrease beyond $R_{\rm m}$, $n_{\rm eq}$($R_{\rm out}$) = 1000 cm$^{-3}$ \\
see text and Fig.\ 7    &  &  & 
+ increase within 2 $\times$ $R_{\rm in}$, by a factor of 2 \\
  &  &  &  \\
\hline
  &  &  &  \\
Relative abundances & constant & $X$(\doce) : $X$(\trece) :
$X$(C$^{17}$O) = 
10$^{-4}$ : 1.5 10$^{-6}$ : 2 10$^{-7}$ & assumed value for $X$(\doce) \\   
 &  &  &  \\
  &  &  &  \\
\hline
\end{tabular}
\begin{tabular}{|l|c|l|}
\hline
 &  & \\
{\bf Other parameters}  & {\bf Values} & comments \\ 
& &  \\
\hline\hline
 &  & \\
Axis inclination from the plane of the sky & 6$^\circ$ & from optical and
CO data  \\
 &  & \\
\hline
 &  & \\
Axis inclination in the plane of the sky (PA) & 85$^\circ$ & from
optical and CO data \\
 &  & \\
\hline
& & \\
Distance  &  300 pc & from luminosity and mass values (Sect.\ 1) \\
 & &  \\
\hline
 & & \\
LSR systemic velocity &  2.5 \kms & from CO data  \\
 & & \\
\hline\hline
\end{tabular}
\end{center}
\end{table*}

\section{Results and conclusions}

Our SMA observations of \doce\ \jdu, \trece\ \jdu, \doce\ \jtd, and
\dsiete\ \jtd\ in \goham\ (Sect.\ 2, Figs.\ 1 to 5) have yielded
high-quality, gorgeous maps. The maps have been satisfactorily
reproduced by our model of disk in keplerian rotation (see Sect. 3,
Fig.\ 6, A.2). From these maps, taking into account our model fitting
of the data and even directly from them (see Sect.\ 3), we have been
able to derive the main structure, dynamics and physical conditions of
the nebula. See parameters of the model and best fitting in Sect.\ 3,
Fig.\ 7, and Table 1. We assume a distance of 300 pc, see Bujarrabal et
al.\ (2008, Paper I) and Sect.\ 1, the dependence of the fitting on the
distance is given in Table 1. See further discussion on the fitting
procedures and accuracy in the Appendix.

   \begin{figure*}
   \centering \rotatebox{270}{\resizebox{8cm}{!}{ 
\includegraphics{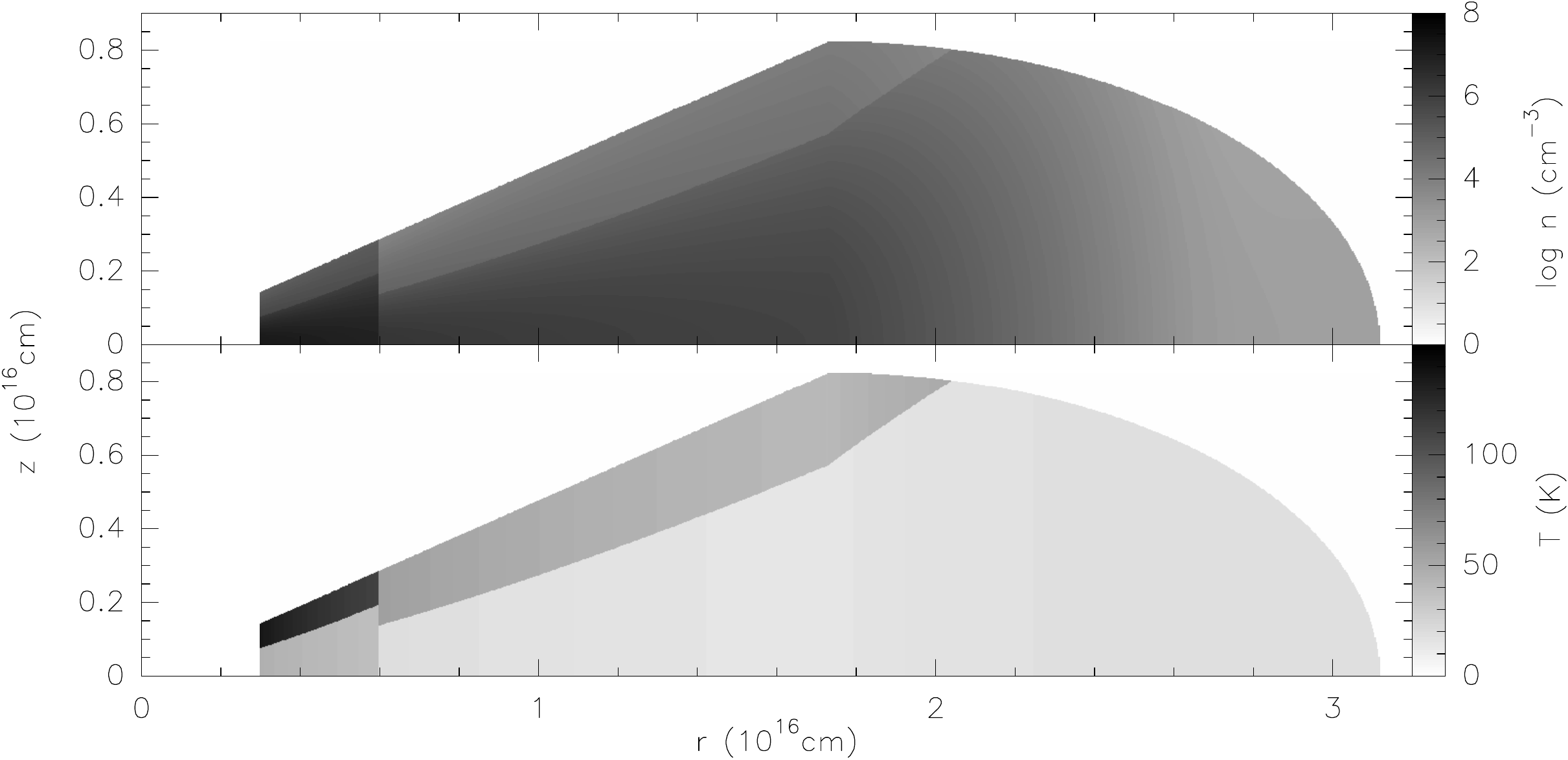}
}}
   \caption{Density and temperature distributions in our best-fit
   model. $r$ is the distance from the rotation axis, and $z$ is the
   distance to the equator.
}
              \label{maps}%
    \end{figure*}

The CO emitting gas in \gohab\ forms a rotating disk that is probably
quite coincident with the whole nebula, since the CO extent is even
larger than the HST image (showing scattered light and absorption by
dust grains, see Figs.\ 1 to 3). We recall, however, that
the disk shape given here is in fact that of the region in which CO is
rich.  Other nebula components in which molecules are severely
underabundant, due to photodissociation by the interstellar or stellar
UV field or to depletion onto grains, are not probed by our maps.

Our data confirm the previous conclusion (Paper I) that the disk is
very nearly in keplerian rotation and shows a clear axial
symmetry. However, we detect, mainly from \trece\ \jdu\ maps (Figs.\ 2,
3) and their comparison with the model predictions, that there is a
condensation in the southern disk that has no counterpart in the rest
of the nebula; see discussion on its properties and meaning in 4.1.

The general dynamics and structure of the nebula are relatively well
determined from the maps. We note that the shape of the model disk
essentially corresponds to the region in which, due to shielding from
the external UV field, CO is still abundant. The shape we deduce is
compatible with the general structure expected for the CO-rich parts of
disks around young stars, see e.g.\ {\citet{jonkheid06,jonkheid07}.

The distributions in the model of the number density ($n$) and
temperature ($T$) require a deeper discussion. The difference between
the distributions of \doce\ and \trece\ \jdu\ emission is particularly
useful for such a purpose. The bright rims, more or less parallel to
the equator, clearly seen in the optically thick \doce\ emission are
not present in \trece. Since \trece\ \jdu\ is certainly much less
opaque than \doce\ \jdu, this indicates a significant increase in
temperature and decrease in density in regions sufficiently separated
from the equator. We note that this decrease of the density with the
distance to the equator, $z$, appears in our model as a result of the
jump in temperature, since the variation of the density with $z$ is
assumed to be just given by the equation of hydrostatics for a rotating
disk. Our model reproduces the observations assuming a characteristic
(sharp) increase of the temperature by a factor 3. We note that
the \trece\ \jdu\ emission comes quite accurately from the opaque
equatorial disk seen in the HST images (Figs.\ 2, 3), while the bright
rims detected in \doce\ \jdu\ at large distances from the equator
correspond to the flaring-disk edge that appears in the visible and
NIR as a result of the scattering of stellar light by dust (Figs.\ 1,
3).

Similar changes in the physical conditions within the disk have been
predicted from theoretical considerations \citep[see
e.g.][]{dullemond07}. Some previous maps of disks around young stars
have also suggested an increase in temperature of this order
\citep[e.g.][]{dartois03}.

Comparison of the \doce\ and \trece\ maps also suggests that $n$
decreases and $T$ increases at large distances to the center in the
plane (large values of $r$), in view of the significantly less extended
emission of \trece\ \jdu\ and the relative maxima found in \doce\ \jdu\
at such large distances. In this case the variations of the temperature
and density are assumed to be independent, indeed we find that the
density decrease factor must be larger. A general description of the
spatial distribution of the physical conditions can be seen in Fig.\
7. We recall that we are assuming that the molecular abundances are
constant within the whole disk, because the existent data do not allow
to consider independently variations of the abundances together with
variations of the density and temperature. Therefore, the decrease in
density for large $r$ could (at least partially) reflect a decrease of
the \trece\ abundance in those regions. (This does not affect the rest
of our conclusions: the temperature must increase for high values of
$z$ and of $r$, to explain the higher \doce\ intensity, and the density
for high $z$ must decrease to satisfy the hydrostatics equations.)

A clear asymmetry is seen in the \doce\ brightness distribution with
respect to the disk equator, i.e\ between the eastern and western parts
of the nebula. This effect is due to self absorption: the cold equator
absorbs the emission coming from the hot, outer regions at high $z$
that are placed behind it (corresponding to the east part of the CO
image), which of course does not affect the emission of the hot layer
that is closer to us. This asymmetry is very well reproduced by our
model calculations, which accurately take into account opacity effects
and radiative interactions between the different parts of the nebula.

We note the very low degree of local turbulence deduced from the 
model fitting, \lsim\ 0.1 \kms. Very small values of the
microturbulence velocity are often found in rotating disks
\citep[e.g.][]{panic08}, even in the rare cases of disks rotating
around evolved stars \citep[][]{bujetal05}.
 
The total central mass responsible for the observed rotation is deduced
to be $\sim$ 3 \ms. We have seen in Sect.\ 1 and Paper I that this
mass value seems somewhat high compared to that derived from the
comparison of the total luminosity with evolutionary tracks, $\sim$ 2
\ms\ (with high uncertainties).  As we discussed in Paper I
and Sect.\ 1, this value may include as much as $\sim$0.5 \ms\ due to
the mass of very dense and very inner regions of the disk (not well
probed by our observations, which do not attain the required very high
resolution). Therefore, the comparison of the mass values derived from
the evolutionary tracks and from the disk dynamics may still imply the
presence of a binary star in \gohab. The conclusion on the binary
nature of the central star depends, unfortunately, on the comparison of
theoretical evolutionary tracks with the uncertain values of the
stellar luminosity and surface temperature, values that cannot at
present be determined with enough accuracy (see Paper I and Sect.\ 1).

Our model also gives the total mass of the (extended) disk. From
integration of the model density, we find a value of $\sim$ 0.01
\ms. As in our previous work (Paper I), we find a disk mass value much
smaller than the total mass derived from dust emission ($\sim$0.3 \ms;
Sects.\ 1, 3). This may be due to a strong gas depletion onto grains,
mainly in the densest regions of the disk 
\citep[][]{dutrey97,thi04,panic08}, which could yield a higher
dust mass and lower molecular abundances than expected. We think that
this effect could at least partially explain the discrepancy found
between the mass derived from dust emission and from CO maps.
Unfortunately, dust continuum emission has been found to come
dominantly from the very center of the disk, which is poorly mapped in
our 1$''$-resolution CO line maps, avoiding a proper comparison of both
distributions. A future study of the effects of depletion on the mass
determinations in this source will require significantly deeper maps of
the continuum emission, tracing dust up to distances comparable to
those observed in CO emission, as well as higher resolution maps
of CO lines, accurately mapping the inner disk regions.

\subsection{The southern condensation}

As we have discussed in Sect.\ 3, there is an asymmetry in our data,
particularly in the \trece\ \jdu\ maps, that cannot be accounted for by
an axially symmetric model (Figs.\ 2, 3). We can see an emission excess
in the southern part of the nebula, at about 1\farcs3 from the
nebula center (6 10$^{15}$ cm for the assumed distance of \gohab) and
between about +4.5 and +5.5 \kms\ LSR, which has no counterpart in the
corresponding northern emission at more negative velocities. This
asymmetry cannot be due to opacity or excitation effects, because it is
clearly more conspicuous in the \trece\ \jdu\ and \dsiete\ \jtd\ lines
than in the \doce\ \jdu\ one, which is more opaque, and than in the
\doce\ \jtd\ line, which requires a higher excitation. The fact that
the intense maximum is more clearly seen in \trece\ \jdu, which is
mostly optically thin, strongly suggests that it is due to a
condensation, i.e.\ to that there is a significant increase of the
density in those regions. We will discuss in this section the
properties and possible origin of this condensation, which could be
protoplanetary.

It is remarkable that such strong asymmetries between regions with
relatively positive and negative velocity are not frequent in CO-rich
rotating disks \citep[][]{simon03,pietu05,simon00}, even in
observations of \trece\ lines. An exception may be the \docho\ maps in
DM Tau by \citet{dartois03}, not their \trece\ data, which present an
emission excess at certain velocities comparable to that in our \trece\
maps. More subtle departures from the keplerain dynamics have been also
identified in AB Aur (as well as in our maps of \gohab; see discussion
below).

We have tried to estimate how much mass is represented by this
excess brightness, although this observational feature is very
difficult to model due to the severe lack of information on its
nature.  A lower limit can be obtained if we assume that the excess
mass is just proportional to the total extra intensity in \trece\
\jdu. The increase of the line intensity at these velocities represents
somewhat less than 20\% of the total (velocity integrated) line
intensity. Since the total mass of our disk model is $\sim$ 0.01 \ms,
we deduce that the mass of the condensation is $\sim$ 1.5 10$^{-3}$
\ms, almost identical to the mass of our planetary system or to the
mass of Jupiter.

We recall that this mass value must be considered as a lower limit. The
excess mass is proportional to the total extra intensity in
\trece\ \jdu\ if this line is completely optically thin and if the
excitation in the clump is the same as in the surrounding gas. We know
that the first assumption is not completely true even for lines of
sight not intersecting the condensation. The second assumption is also
improbable, since the density increase must be important and since the
opaque \doce\ \jdu\ line also presents some intensity increase in these
regions, suggesting that they are hotter than the corresponding
northern regions. The presence of some opacity also in \trece\ \jdu\
will lead to a perhaps important underestimate of the mass derived 
from our simple estimate. The effects of the different excitation
should be less important. A higher excitation increases the partition
function of the molecule and decreases the level population difference,
which leads to lower emission in optically thin lines (although, of
course, a higher temperature means a higher intensity for high
opacities). Therefore, both assumptions will in general lead to 
underestimates of the clump mass.

We have performed a better estimate introducing a density increase in
some parts of our model disk (destroying the axial symmetry).  We have
estimated, first, that to model the asymmetry in the \doce\ \jdu\ maps
we need an excess temperature by a factor 1.5 in a southern
disk region closer than 10$^{16}$ cm. Then, we have tried to roughly
fit the \trece\ \jdu\ data increasing the density by some factor, with
respect to standard law (derived from our fitting of the whole
  data, Sect.\ 3, Table 1).
A significant factor ($\sim$10) is found to be
necessary to account for the dilution in the beam and some incipient
optical depth in this line. The model predictions for the southern
brightness increase in \dsiete\ \jtd\ are also compatible with the
observations. See a detailed discussion on our tentative
modeling of this clump in Sect.\ A.3. With these conditions, we
find that the condensation contains a total mass $\sim$ 6 10$^{-3}$
\ms, a few times our planetary system.

We note that these estimates do not exclude the presence of more
compact components, not detectable with our angular resolution. We also
note that a proto solar system must contain a significantly higher mass
than the final planetary system, since significant gas ejection is
expected before planets form from such very premature condensations. We
also recall that our poor knowledge of the physical properties of these
clumps and the poor observational information on it, which practically
remains unresolved in our data, prevent any attempt of detailed
modeling of the data.

   \begin{figure}
   \vspace{-2.7cm}
\hspace{-0.2cm}
\rotatebox{0}{\resizebox{9cm}{!}{ 
\includegraphics{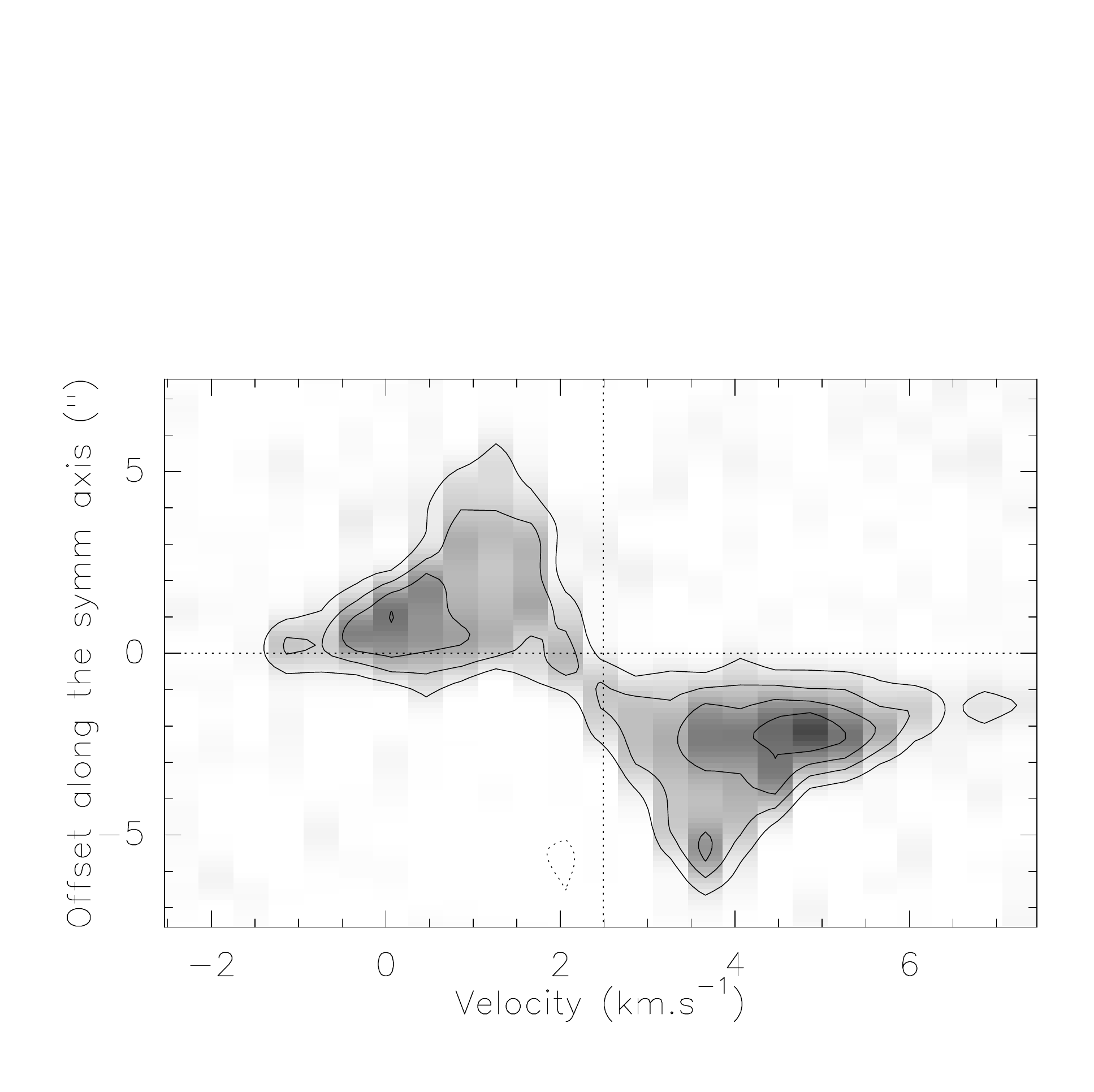}
}}
   \vspace{-0.3cm}
   \caption{Position-velocity diagram of the \trece\ \jdu\ emission
   from \gohab, obtained along the disk direction
   (P.A.=--5$^\circ$). The southern emission excess can be seen at
   positive velocities. 
}
              \label{maps}%
    \end{figure}

On the other hand, we note that our fitting of the \trece\ emission at
these velocities is only accurate for the intensity of the clump. 
In the observations, the brightness maximum appears displaced outwards
by 0.5 arcsecond, with respect to the position expected from our
fitting (see Appendix). This effect cannot be avoided assuming that the
condensation is farther from the star, since then the velocity field
we are assuming (the same keplerian law deduced from the fitting of the
overall dynamics, Sect.\ 3) would imply that its emission occurs at too
low relative velocities. This displacement can easily be seen in the
observations: note the anomalously long distance from the star of the
maximum in \trece\ \jdu\ emission at about 5 \kms, compared to the
brightness distributions at about 3.5 \kms. 
Therefore, the presence of
a disturbance in the dynamics is necessary to explain the
observed properties of the southern condensation, both the disturbance
in the dynamics and the condensation being very probably associated.

The standard conditions theoretically required to produce gravitational
instabilities in the disk \citep[e.g.][]{durisen07} are not satisfied
for this condensation. In \gohab, instabilities would only
occur at a less than $\sim$10$^{15}$ cm, the distances at which
theoretical models tend indeed to predict the presence of
protoplanets. The size deduced from our data for the condensation is
also too large compared to the Hill sphere, which defines the
gravitational domain of a protoplanet \citep[e.g.][]{lissauer07}, if we
do not assume the presence of undetected compact components. In fact,
the mass of the central component required by the large extent of the
condensation would be $\sim$ 0.5 -- 1 \ms, i.e.\ that of a low-mass
protostar.

From the observational point of view, however, there exists other
evidence of the presence of important condensations, probably due to
instabilities in the disk, at similar large distances from the
star. For instance, the well known spiral structure in AB Aur appears
at a distance smaller than about 100 AU or 1.5 10$^{15}$ cm
\citep{oppenheimer08, millan06}, but spiral-like dust condensations are
also detected at several hundred AU \citep[][]{fukagawa04} and seem to
have some counterpart in CO emission
\citep{pietu05,lin06}. \citet{pietu05} and \citet{lin06} also detected
departures from the general keplerian velocity field in the spiral arms
of AB Aur, comparable in magnitude to those found by us in \gohab. 
To better compare our observations with previous results, we represent
in Fig.\ 8 the position-velocity diagram along the disk direction in
\gohab, obtained from our \trece\ \jdu\ data, in which the excess
emission is clearly seen at positive velocities. The p-v diagram in
\gohab\ is found to be quite comparable to that obtained by
\citet{lin06} in AB Aur, in spite of the different angular and velocity
scales. The deformation of the dynamics in AB Aur associated with
the spiral instability therefore results in a velocity pattern quite
similar to that found in \goham. On the other hand,
\citet{lafreniere08} have reported the detection, from direct imaging
and spectroscopy, of a planet with a mass of about 8 times that of
Jupiter at 5 10$^{15}$ cm from a young star, although systems involving
such long distances seem
to be rare \citep{lafreniere07}. Planets with a few times
the mass of Jupiter have been also found around the A-type stars
Fomalhaut and HR\,8799 \citep{kalas08,marois08}, at distances long,
$\sim$ 100 AU, though somewhat shorter than that of our condensation.
Finally we note the recent detection, from dust emission mapping at
1.3cm wavelength, of a condensation containing 14 times the mass of
Jupiter at 65 AU from HL Tau, which was interpreted to be
protoplanetary \citep{greaves08}.

The southern condensation could also be associated with the
presence of a low-mass stellar or quasi-stellar companion, which is
able to gravitationally capture a sizeable fraction of the overall disk
material and may explain the anomalous dynamics found in the
condensation. The contribution of the companion mass (\lsim 1 \ms)
to the total central mass could be necessary to explain that the mass
value derived from the overall disk rotation (i.e.\ derived from data
on quite extended regions and regardless of the peculiar kinematics of
the condensation) is perhaps too high for the observed total
luminosity; see more details above.

In summary, we have identified a condensation in the southern part of
the disk, at about 1\farcs 3 from the disk center, that probably
contains a mass of a few times that of Jupiter. This condensation could
be protoplanetary, in a very preliminary evolutionary stage; we have
noted however that theory predicts that planets tend to form at much
smaller distances from the central star, although some observational
results support the formation of condensations at such a long
distance. This clump could also be due to material gravitationally
captured by a stellar or quasi-stellar companion, whose presence has
been independently proposed to explain the total central mass derived
from the disk rotation velocity. We recall that the condensation seems
to be associated with a significant distortion of the general keplerian
velocity field of the bulk of the nebula; this distortion could be
explained both if the clump formation is due to the presence of a
low-mass stellar companion or to gravitational instabilities within the
disk.



\bigskip
\begin{acknowledgements}
We are grateful to Asunci\'on Fuente and Mario Tafalla for very
helpful comments on the manuscript.    
\end{acknowledgements}

\bibliographystyle{aa}
\bibliography{../protostars}

\appendix

\section{Selection of the best fitting and accuracy of the parameter values}

   \begin{figure*}
   \centering \rotatebox{270}{\resizebox{8cm}{!}{ 
\includegraphics{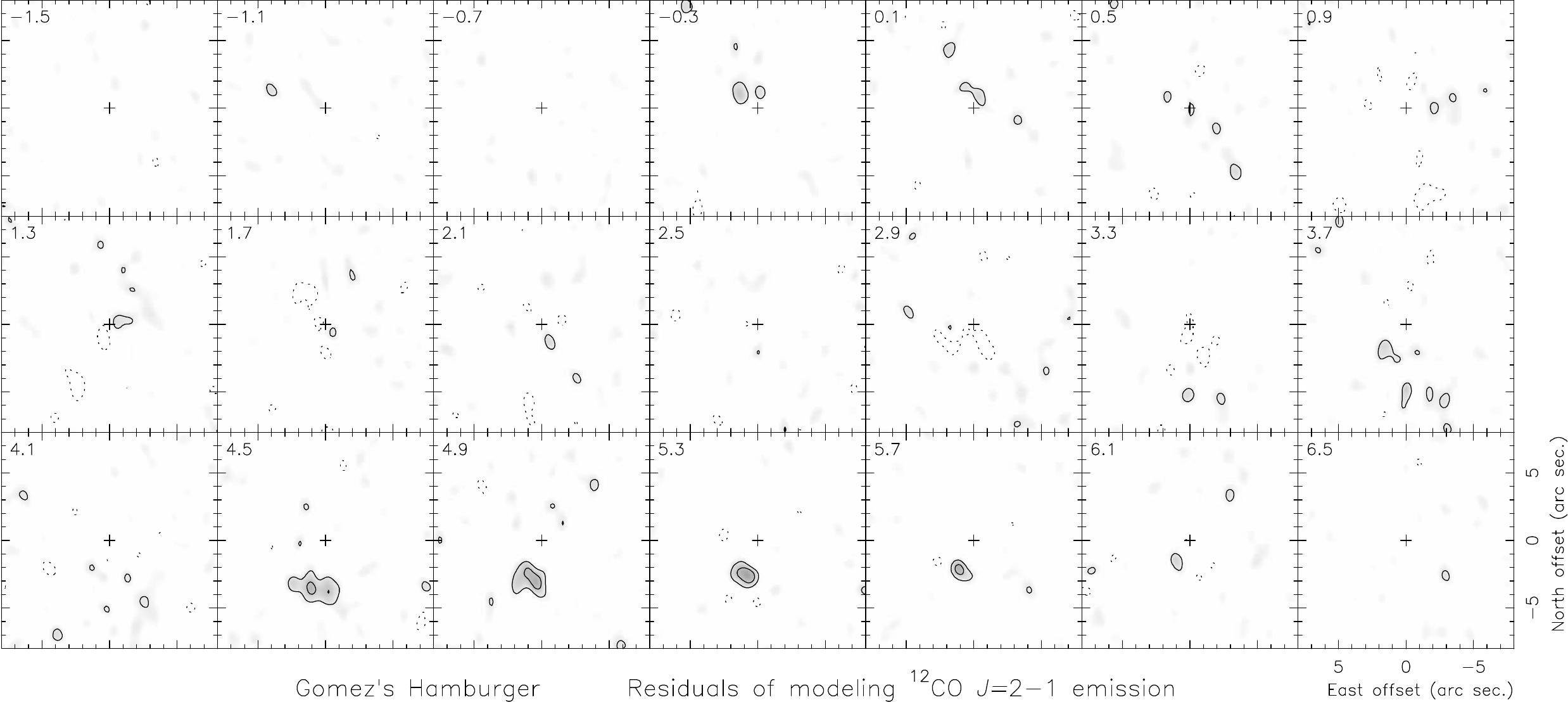}
}}
   \caption{Residuals (observations minus synthetic maps) of the
   \doceCO\ \jdu\ line brightness from our best model fitting for the
   \gohab\ disk. The spatial scale and contours are the same as in 
   the observations and predictions, Figs.\ 1, 6.  }
              \label{maps}%
    \end{figure*}

\subsection{Criteria for acceptable models}

The general criterion we have chosen to select acceptable models is the
comparison of the predicted images with the observed ones. Some authors
\citep[e.g.][]{dutrey07,pety06,isella07} perform such a comparison in
the Fourier transformed plane of the visibilities. The selected model
parameters are then those yielding the smallest residuals, after
considering `blind' variations of the parameter values. Their method
has the advantage that is very objective and that uncertainties
introduced by the cleaning process are avoided.
Other authors
\citep{mannings97,mannings00,fuente06}, follow however a more intuitive
approach, directly comparing the images.

   \begin{figure*}[t]
   \centering \rotatebox{270}{\resizebox{8cm}{!}{ 
\includegraphics{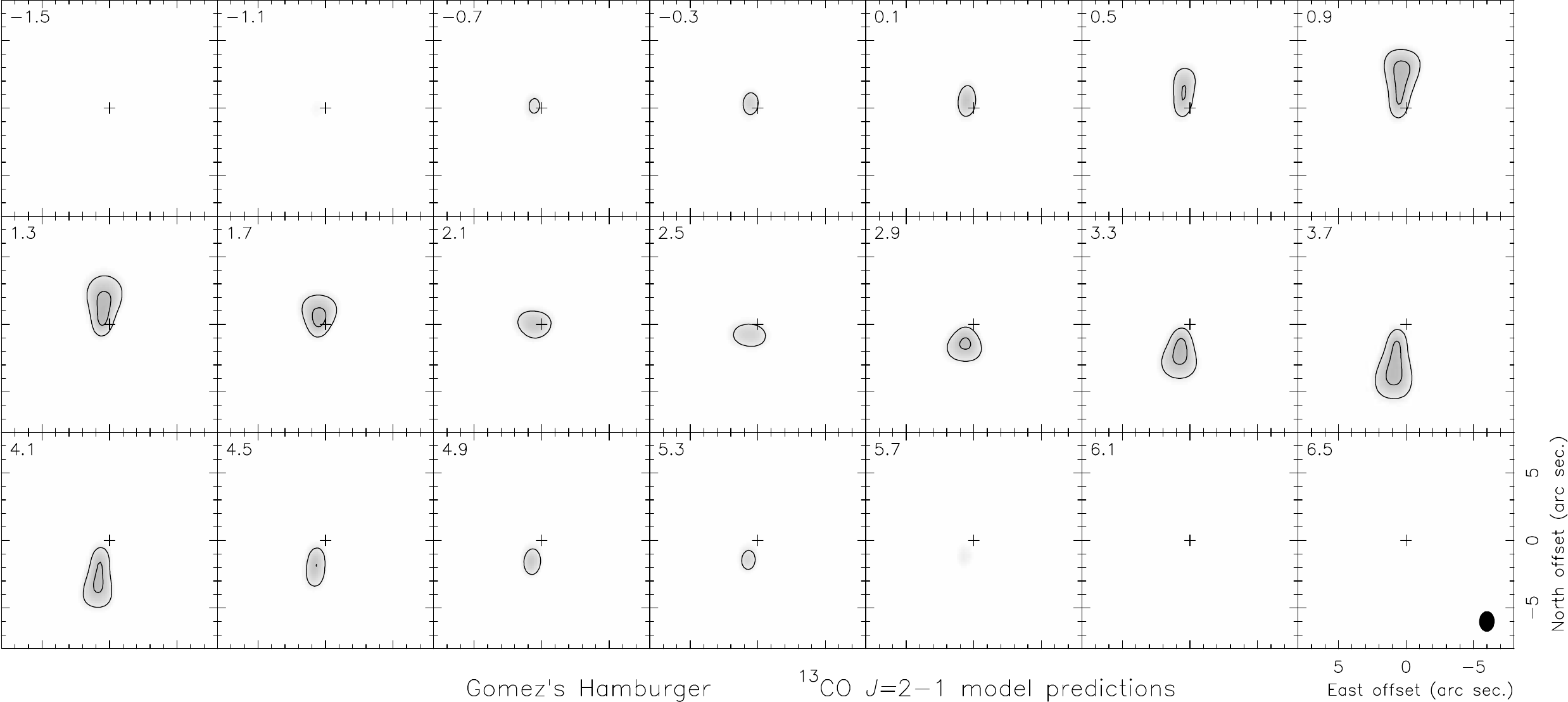}
}}
   \caption{Predictions of the \treceCO\ \jdu\ line brightness from our
   best model fitting for the \gohab\ disk. The spatial scale and 
   contours are the same as in the observations, Fig.\ 2.
}
              \label{maps}%
    \end{figure*}

   \begin{figure*}
   \centering \rotatebox{270}{\resizebox{8cm}{!}{ 
\includegraphics{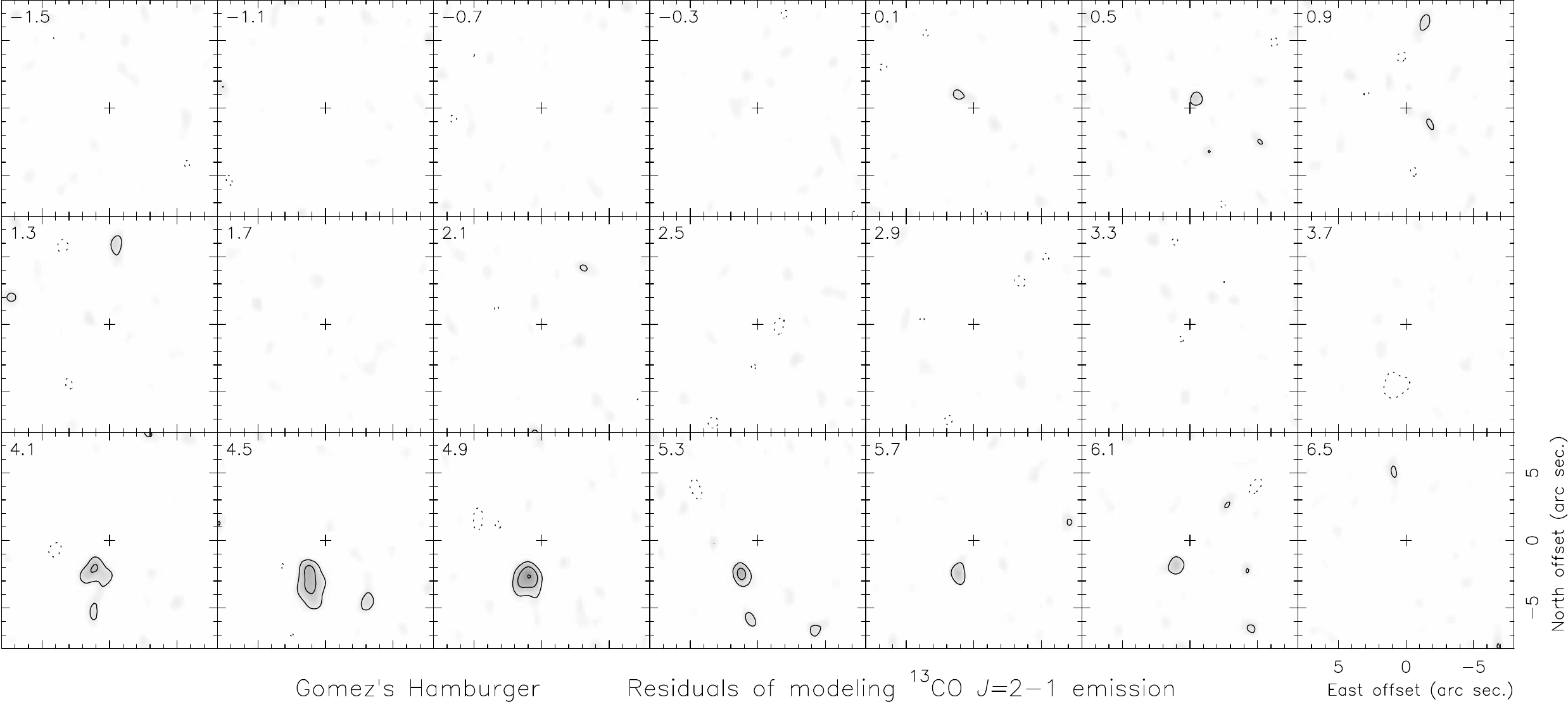}
}}
   \caption{Residuals (observations minus synthetic maps) of the
   \treceCO\ \jdu\ line brightness from our best model fitting for the
   \gohab\ disk. The spatial scale and contours are the same as in 
the observations and predictions, Figs.\ 2, A.2.  }
              \label{maps}%
    \end{figure*}

In our case, the very high number of free parameters prevents a blind
analysis of the resulting residuals for any combination of parameter
values, which would imply an exceedingly high amount of
calculations. Even the definition of `free parameter' in such a complex
structure is difficult, and it is not clear at what extent we can vary
all the properties of the disk. Even if theoretical constraints to our
model parameters have been widely taken into account (see Sect.\
3),we have more than 15 (more or less free) model parameters: 5
defining the geometry, 2 for the dynamics, 5
for the temperature, 3
for the density, plus finally the $^{12}$CO/$^{13}$CO/C$^{17}$O
abundance ratio. It is difficult to decrease such a large number of
parameters in view of the different components identified from
the maps: the hotter and less dense fringes separated from the equator,
an outer less dense region, the cold central part of the disk, and the
central hotter region.
 
Our procedure, on the other hand, allows an intuitive analysis of the
relevance of the different parameters. We can so deduce that some
parameters are not relevant to fit some observational properties, and
that certain observational features are related to just one or two
model parameters. For instance, the total size of the disk is selected
to match the total extent of the image, which depends weakly on e.g.\
the rotational velocity.

It has been argued that in sources comparable in angular size with the
telescope resolution the comparison performed in the image plane is not
accurate. This is not our case, since \gohab\ occupies about 30 square
arcseconds, a surface almost 20 times larger than the beam of our
\doceCO\ \jdu\ data. The cleaning process always introduces noise, but,
besides numerical noise, it is mostly due to uncertainties in the
calibration of the complex visibilities, i.e.\ to that the actual beam
shape is not exactly equal to the theoretical beam function (for the
$uv$ coverage of the observation). So, subtraction of the convolved
`clean components' and the subsequent convolution to get the final
image introduces spurious features, mainly when the $uv$ sampling is
poor. However, such unavoidable uncertainties in the amplitude
and phase calibration appear also if the fitting is performed in the
$uv$ plane, and the limitation in the dynamic range of the observations
due to them applies to both the sky plane and the Fourier
transformed plane. On the other hand, we also note that Fourier
transformation of the predicted brightness (which must be calculated
for a finite grid of points in space coordinates from the standard
radiative transfer equation) also introduces numerical noise.

As we will see below in actual cases, it is practically impossible to
define an absolute best fitting model in our case, because of the
complex data and model. Instead, we will adopt criteria to consider
that a set of parameter values is acceptable and, from cases in which
they are not satisfied, will give limits to these values. This process
will only be detailed for the most representative parameters: radius
and width of the disk, central mass (i.e., rotational velocity law),
typical densities and temperatures, ... Among the acceptable models
there is no significant difference, both in predictions and in values
of the relevant disk properties; one of them has been chosen as our
best fit.

To select the acceptable models, we will basically use the comparison
of the predictions with the observations of the \doceCO\ and \treceCO\
\jdu\ lines, in which the S/N ratio and the spatial and spectral
resolutions are particularly high. We have checked that the predictions
for the \jtd\ lines are also satisfactory. The criteria used to select
acceptable models are:

\noindent 
{\bf c1)} The differential image, i.e.\ the observed minus synthetic
velocity channels, must show in the regions of each channel where
emission is present an {\em rms} noise not exceeding $\sim$ 1.5 times
that present in adjacent regions with no emission. Those regions, with
noise not larger than 1.5 times that found in adjacent ones, are only
slightly noticeable in the differential maps.

\noindent
{\bf c2)} In the differential \jdu\ image, the residuals must be
smaller than 2 times the spurious contours (due to noise) seen in
adjacent regions ($\sim$ 0.6 Jy/beam, two contours, in our \doce\ \jdu\
images), and no residual \gsim 0.3 Jy/beam must appear systematically,
i.e.\ in the same spatial offsets for several velocity channels. We
note that in the observed \doce\ \jdu\ maps one can identify noise
features more intense than 0.3 Jy/beam, so residuals not exceeding the
above limits are again not clearly different than the observation
noise.

These conditions are relaxed for velocities around 4.5--5.5 \kms\ LSR,
which present a strong emission clump with no counterpart in the
equivalent blueshifted emission.
This excess cannot be due to opacity or excitation effects, since it is
more prominent in the \treceCO\ \jdu\ maps than in the \doceCO\ \jdu\
ones and much more than for \doceCO\ \jtd. The excess is also clear in
the \dsiete\ \jtd\ maps. The fact that this excess is so remarkable in
the less optically thick \trece\ emission strongly suggests that is
mainly due to the presence of a gas condensation, rotating at about 2.5
\kms\ at a distance of about 5 10$^{15}$ cm from the star. Our model
shows axial symmetry and therefore cannot explain this excess; we have
chosen to mainly fit the emission from the rest of the nebula. A
tentative model fitting of this feature is presented in A.3, and the
consequences and possible origins of the presence of such a
condensation are discussed in Sect.\ 4.1.

In figure A.1, we present the residuals of our fitting for the \doce\
\jdu\ maps. Predictions and residuals for \trece\ \jdu\ are shown in
Figs. A.2 and A.3. Except for the velocities around 5 \kms\ the
typical rms noise out of the emitting regions is of about 0.1 - 0.11
Jy/beam (slightly less for \trece\ \jdu), it does not exceed 1.5 - 1.7
Jy/beam in the regions presenting emission. We see that the largest
residuals in the differential images do not reach two contours (noise
reaching one level is also seen out of the emitting regions).

\begin{table*}[]
\caption{Relative range of acceptable values, around those given in
  Table 1, of the main parameters
  defining the model disk in
  Gomez's Hamburger.}
\begin{center}                                          
\begin{tabular}{|l|c|l|}
\hline\hline
& & \\
{\bf Representative parameter}  & {\bf acceptable range} &  comments \\ 
& & \\
\hline\hline
   &  &  \\
Disk radius & $\times$/$\div$ 1.1 & $R_{\rm out}$ and $R_{\rm m}$  \\
   &  &  \\
\hline
   &  &  \\
Disk width  & $\times/\div$ 1.1 & $H$  \\
   &  &  \\
\hline
   &  &  \\
Keplerian velocity & $\times/\div$ 1.1 & $V_{\rm t}$ \\
   &  &  \\
\hline
   &  &  \\
Temperature     & $\times/\div$ 1.1 & characteristic value   \\
High-$z$ jump   & between 2 and 5 & depends on calibration, possible
exc.\ effects (see A.2) \\
   &  &  \\
\hline
   &  &  \\
Density  & $\times/\div$ 1.5 & characteristic value   \\
Outer disk density & $\times/\div$ 3 & only given by \doce\ data,
possible exc.\ effects (A.2) \\
   &  &  \\
\hline
   &  &  \\
Turbulent velocity  & $\leq$ 0.1 \kms\ & local velocity dispersion  \\
   &  &  \\
\hline
   &  &  \\
$X$(\doce)/$X$(\trece) rel.\ abundance  & $\times/\div$ 1.5 &  \\
$X$(\doce)/$X$(C$^{17}$O) rel.\ abundance  & $\times/\div$ 2 &  \\
   &  &  \\
\hline\hline
\end{tabular}
\end{center}
\end{table*}

Of course, between 4.5 and 5.5 \kms\ the situation is worse, reflecting
the asymmetry in the disk density mentioned above. We can notice some
other minor features in the images that are not accounted for by our
model: For instance, the \trece\ line emission at 3.7 \kms\ is less
extended than expected, which is not the case at 4.1 \kms. We also note
in our \trece\ maps a weak extent towards the north at about 1.3 \kms,
but practically at the noise level. In \doce\ \jdu\ we see a similar
protrusion, but not exactly in the same position. We finally note in
some observed panels that the emission from regions close to the star
(see e.g.\ the \trece\ \jdu\ emission at $\sim$ 0.5 \kms\ and the
\doce\ emission at 3.7 \kms) is somewhat wider than the predictions. The
width of our disk is essentially given by the equation of hydrostatics
(Sect.\ 3), under the standard theory of massive flaring disks; it is
possible that these usual theoretical requirements are not fully
satisfied in the innermost regions of the disks, further analysis of
this phenomenon obviously requires higher-quality observations.

\subsection{Uncertainty in the fitted parameters}

We have estimated the uncertainties in derived values of the model
parameters by varying the values for each one (while the others remain
unchanged) and checking for which values the above conditions, {\bf c1,
  c2}, are clearly not satisfied. This has been only done for the main,
most representative parameters, for instance the radius disk $R_{\rm
  out}$, the characteristic density, etc. The results are
summarized in Table 1. We also present below some cases in which
the uncertainty in the derived parameters requires some discussion. 

We have not tried to fully consider the parameter uncertainties when
two or more parameters are allowed to vary. For example, the density
range is slightly larger than that given in the table if we allow the
temperature also to vary, since both variations are in some way 
compensated. In general, however, the ranges do not differ very much
from our standard uncertainty brackets. The density is
mostly fixed by the emission of \trece\ and that of \doce\
\jdu\ from outer regions, while the temperature law is
mainly given by the \doce\ \jdu\ and \jtd\ maps.

We note the uncertain determination of the density in the outer disk
($r$ $\sim$ $R_{\rm out}$), which is only given by the \doce\ emission,
since \trece\ is not detected in this region; see Table A.1. We also
recall that the assumption of level population thermalization may not
be valid for these diffuse regions, which may imply that the density in
them is somewhat larger than the values given here, perhaps closer to
3 10$^{3}$ cm$^{-3}$. 

Important problems with the thermalization assumption are not expected
for the relatively high densities of the rest of the nebula and the
analyzed transitions (Sect.\ 3). The \jsc\ transition is significantly
more sensitive to these effects, due to the relatively high Einstein
coefficients of high-$J$ transitions. We also expect underpopulation of
the $J$=6 and $J$=5 levels in the high-$z$ low-density regions, with
$n$ $\sim$ 10$^{4}$ cm$^{-3}$, leading to 6--5 brightness temperatures
under 10 K. This line would mainly come from this high-temperature
surface, and we think that this phenomenon is in fact the responsible
for the non-detection of \doce\ \jsc\ in our observations. Any further
discussion is not justified in view of the lack of accurate information
on the \jsc\ emission. This relative underpopulation of high-$J$ levels
in the layers at high absolute values of $z$ may lead to a relative
overpopulation of the $J$=2 level, and therefore to more emission than
expected in \jdu. This effect could lead to slightly smaller jumps of
the temperature, perhaps closer to factor 2.

   \begin{figure*}
   \centering \rotatebox{270}{\resizebox{8cm}{!}{ 
\includegraphics{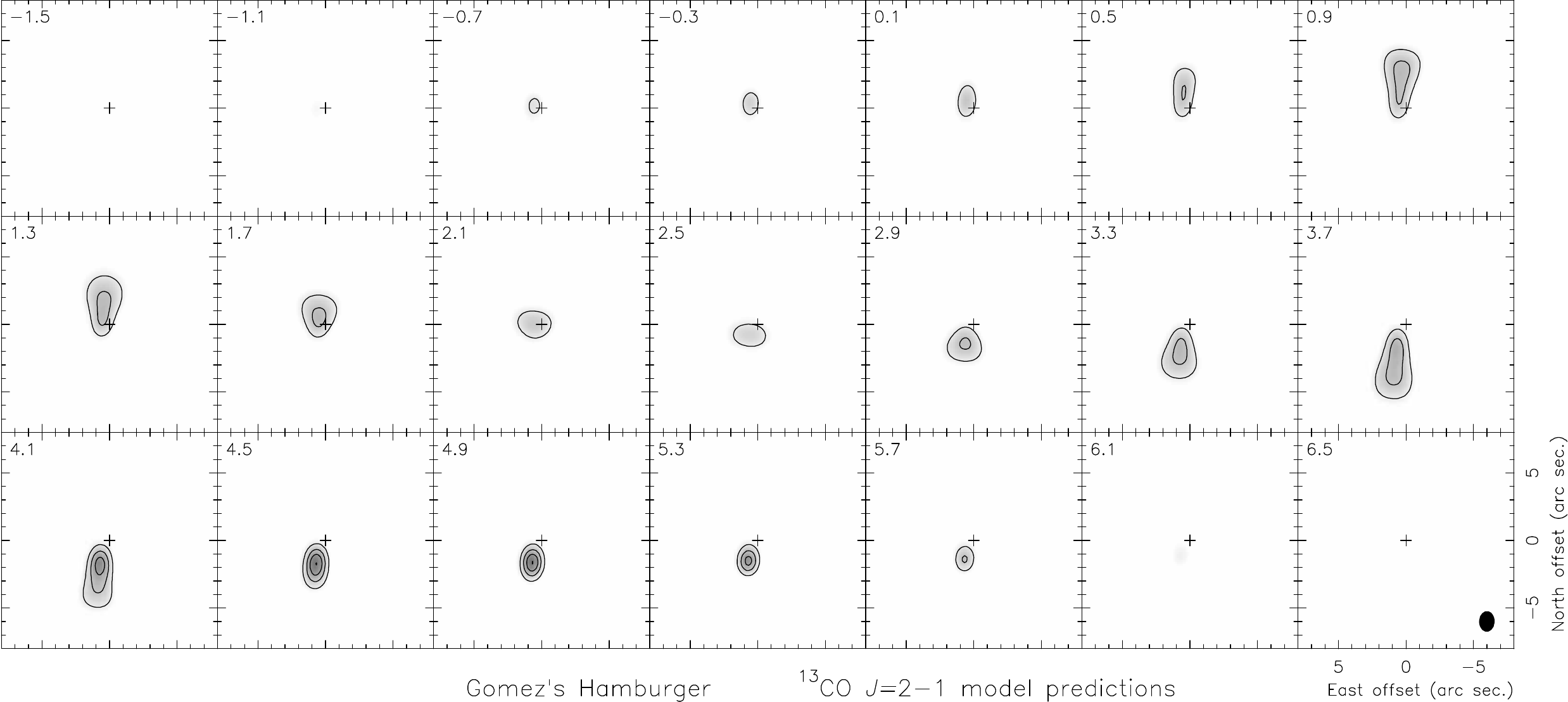}
}}
   \caption{Predictions of the \treceCO\ \jdu\ line brightness from our
   best model fitting for the \gohab\ disk, including the increase in
   temperature and density in a southern clump, as described in App.\
   A.3. The spatial scale and contours are the same as in the
   observations, Fig.\ 2, and our standard model predictions, Fig.\
   A.2.
  }
              \label{maps}%
    \end{figure*}

Some parameter pairs are quite dependent one on the other. This is in
particular the case of the density and the CO total abundance,
since we are assuming LTE and the line opacity depends on the product
of the density and the relative abundance. So both parameters can
freely vary, provided that their product is kept constant. The
indetermination is solved assuming a relative \doce\ abundance of
10$^{-4}$. This is also the case of parameters that we have not
considered separately in this uncertainty analysis, like the
temperature at a given point, $T(R_0)$, and the slope of the
temperature law, $\alpha_T$; instead, as mentioned, we just discuss
the uncertainty in the characteristic temperature. Finally, we note the
case of the rotation velocity and the conditions in the hot and dense
center of the disk. A strong emission from these regions may mimic the
emission of a faster rotating disk in the extreme velocity
channels. So, the rotation velocity is mainly determined from the
emission extent in the channels at moderate velocities, the emission at
the extreme channels depending on both the keplerian velocity and the
emissivity of the central regions. In general, the uncertainty of the
parameters defining the inner, denser region ($r$ $<$ 2 $\times$
$R_{\rm in}$) is high, because this clump is not resolved and the
number of independent observational constraints is small (as also
concluded in Paper I). We can say that an inner region with higher
density and temperature is necessary to attain our strong requirements
to the fitting quality and that it must be smaller than about 10$^{16}$
cm; but we cannot give details on this region.

The outer hot region, for high values of $z$, is hardly resolved in our
\doce\ maps. Therefore, we could also fit our data assuming that this
region is significantly thinner and brighter (in general, hotter) than
in our standard model. We think however that those disk models are less
probable than our standard ones, because our standard jump in
temperature is already quite high (see Sect.\ 3). Moreover, for very
high temperatures in the high-$z$ rim, 
we should also increase significantly the typical density (to
fit \doce\ \jdu, which, for the assumed dependence of the density
with $z$, is only moderately opaque in the high-$z$ regions). This
would then imply too low values of $X$(\trece), to fit \trece\ \jdu,
leading to improbably high values of $X$(\doce)/$X$(\trece),
significantly over 100. In any case, we note that the properties of
this high-$z$ bright rim vary only moderately with its temperature
(because of the moderate opacity in \doce\ \jdu) and depend on the
relative calibration of the \jdu\ and \jtd\ lines; allowing 15\%
variations in the relative calibrations, we can fit the data with
temperature jumps ranging between 2 and 5 (and hot-layer widths not
very different form our standard value).

\subsection{Model (tentative) fitting of the southern brightness maximum}

We have mentioned that there is a relative maximum in the southern part
of the disk that has no counterpart in the north. This maximum is seen
in all our maps, but it is more prominent in \trece\ emission, which is
mostly optically thin, as well as in the \dsiete\ line. Therefore, this
brightness maximum must be mainly due to an increase of the density in
some southern regions, though some increase in temperature is also
necessary to explain the excess observed in \doce\ \jdu.

In our standard model analysis, we have mostly tried to fit the
emission from the rest of the nebula (Sects.\ 3, A1, A2). Any 
attempt to fit the emission excess from this southern clump is very
uncertain, because of the lack of previous experience trying to study
condensations of this kind, including the lack of theoretical modeling,
and because of the poor information contained in our observations,
which scarcely resolve its extent.

Nevertheless, we note that this emission excess can be detected over a
remarkable range of velocities, between about 4.1 and 5.7 \kms\
LSR. This means that the emitting condensation cannot be very small,
the projected velocity dispersion being due to gas emitting from
different distances from the star or from regions showing a significant
variation of the projection of the velocity on the line of sight. In
both cases, we expect typical sizes $\sim$ 5 10$^{15}$ cm ($\sim$ 1$''$
for the adopted distance). We can also assume that the condensation
takes place in the equatorial disk regions from which \trece\ \jdu\
emission comes, because the maximum is so prominent in this line. We
have assumed that the emission comes from region defined by $R_{\rm
in}$ $<$ $r$ $<$ 10$^{16}$ cm, and $z_s$/$r$ $<$ 0.6, where $z_s$ is
the distance between a given point and the plane containing the star
and perpendicular to the equator that gives the extreme projections for
the rotation velocity. We assume that the temperature and density of
this region vary with respect to the standard laws assumed for the rest
of the nebula.

We have estimated that, to explain the (moderate) emission excess found
in \doce\ \jdu, we must assume an increase in temperature in the
southern condensation by a factor $\sim$ 1.5, with respect to our
standard laws for nearby regions. Finally, we have estimated the
excess density in the condensation by comparison of the model
results with the intensity of the \trece\ \jdu. Since the \trece\ \jdu\
emission from these inner regions is not fully optically thin, a
significant density increase, by a factor 10, is necessary. A total
mass increase of about 6 10$^{-3}$ \ms\ is deduced. 

The resulting brightness distribution is shown in Fig.\ A.4.  We can
see that the brightness excess of \trece\ \jdu\ in this southern clump
is reasonably well reproduced, but the location of the predicted
maximum is slightly closer to the star than the observed one. This
cannot be avoided assuming a longer distance from the star for the
clump, because then the velocity of the feature would be less
positive. The velocity field in this region must then be also
disturbed.  See discussion on the interpretation of this emission
excess in Sect.\ 4.1.

\end{document}